\documentclass[a4paper]{article}
\usepackage{RR}
\usepackage{hyperref}
\RRdate{December 2008}

\usepackage{epsfig,psfrag}
\usepackage{amsmath,amssymb,graphics, amscd }
\usepackage{color}
\usepackage{graphicx,psfrag}
\def\C{\mathcal{C}}

\def\H{\mathcal{H}}
\def\F{\mathcal{F}}
\def \R {\mathbb{R}}

\def \S {\mathbb{S}}
\def \ds {\displaystyle}
\def \ind {\mathbf{1}}
\newenvironment{proof}{\noindent  {\bf{Proof}} {\bf:}  \newline}
{\hfill \nopagebreak $\Box$ \newline }

\catcode`é=\active\def é{\'e} \catcode`è=\active\def è{\`e}
\catcode`ë=\active\def ë{\"e} \catcode`ê=\active\def ê{\^e}
\catcode`î=\active\def î{\^\i} \catcode`ï=\active\def ï{\"\i}
\catcode`à=\active\def à{\`a} \catcode`û=\active\def û{\^u}
\catcode`Ž=\active\def Ž{\'e} 
\catcode`=\active\def {\`e} 
\catcode`ˆ=\active\def ˆ{\`a} 
\catcode`=\active\def {\^e} 

\begingroup

\newtheorem{lemma}{Lemma}
\newtheorem{theorem}{Theorem}
\newtheorem{proposition}{Proposition}
\newtheorem{definition}{Definition}

\endgroup

\newcommand {\mm}[1] {\ifmmode{#1}\else{\mbox{\(#1\)}}\fi}

\newcommand{\BB}            {\mm{\mathbb B}}

\newcommand{\e}             {\mm{\varepsilon}}
\newcommand{\rr}            {\sc{r}}

\def \reach {\rm reach}
\newcommand{\NearestSet}          {\Gamma}

\RRauthor{
Fr\'ed\'eric Chazal\thanks{INRIA Saclay, Parc Club Orsay Universit\'e,
2-4, rue Jacques Monod, 91893 Orsay Cedex France, frederic.chazal@inria.fr
}%
  \and
David Cohen-Steiner\thanks{INRIA, BP 93, 06902 Sophia Antipolis, France 
, david.cohen-steiner@sophia.inria.fr}%
\and
Andr\'e Lieutier\thanks{Universit\'e de Grenoble \& CNRS, Laboratoire Jean Kuntzmann, Grenoble, France
}
\and
Boris Thibert\thanks{Universit\'e de Grenoble \& CNRS, Laboratoire Jean Kuntzmann, Grenoble, France,
Boris.Thibert@imag.fr}
}
\authorhead{Chazal \& Cohen-Steiner \& Lieutier \& Thibert}
\RRtitle{Stabilit\'e des Mesures de Courbures}
\RRetitle{Stability of Curvature Measures}
\titlehead{Stability of Curvature Measures}
\RRresume{Ce travail \'etudie l'estimation de la courbure d'ensembles compacts \'echantillonn\'es. Nous montrons que les mesures de courbures 
moyenne, de Gauss et anisotropes des offsets d'un compact $K$ suffisamment ``r\'egulier'', i.e. \`a $\mu$-reach strictement positif, sont stables : elles peuvent \^etre estim\'ees par les mesures de courbures des offsets de tout compact $K'$ suffisamment proche de $K$ pour la distance de Hausdorff.
Nous explicitons le calcul de ces mesures de courbures dans le cas ou $K'$ est un nuage de points dans $\R^3$. Ces r\'esultats fournissent en particulier un cadre permettant de d\'efinir formellement des notions de courbures robustes pouvant \^etre effectivement calcul\'ees.
}
\RRabstract{We address the problem of curvature estimation from sampled compact sets. The 
main contribution is a stability result: we show that the gaussian, mean 
or anisotropic curvature measures of the offset of a compact set K with 
positive $\mu$-reach can be estimated by the same curvature measures of the 
offset of a compact set K' close to K in the Hausdorff sense.
We show how these curvature measures can be computed for finite unions of balls.
The curvature measures of the offset of a compact set with positive $\mu$-reach can 
thus be approximated by the curvature measures of the offset of a 
point-cloud sample. These results can also be interpreted as a framework for
 an effective and robust notion of curvature.}
\RRmotcle{mesures de courbure, cycle normal, inf\'erence g\'eom\'etrique, $\mu$-reach, fonction distance, nuage de points}
\RRkeyword{curvature measures, normal cycle, geometric inference, $\mu$-reach, distance function, point-cloud}
 \RRprojet{Geometrica}  
 \RRtheme{\THNum} 
\RCSaclay 

\begin{document}

\RRNo{6756} 
\makeRR   

\section{Introduction}
\paragraph{Motivation} We present in this work a {\em  stable}  notion of curvature.
A common definition of curvature considers quantities defined pointwisely on a twice differentiable manifold. However, the objects we have to deal with in practice are not twice differentiable: consider the situation where a physical object is known through a sufficiently dense point cloud measured on the object boundary. Intuitively, it seems feasible to infer some meaningful information on the curvature of the physical object itself. Let us assume that we know both the {\em accuracy} of the measure, which is  an upper bound on the distances between the measured points and their closest points on the physical object boundary, and the { \em sampling density}, which is an upper bound on the distance between the points on the object boundary and their closest measured sample points. The fact that the measure  accuracy and sampling density are below a known small value $\epsilon$ can be expressed by saying that the Hausdorff distance (see the definition in  Section \ref{section-background}) between the assumed physical object and the measured point cloud is less than  $\epsilon$.
Even with the guarantee of a small Hausdorff distance, the knowledge of the measured point cloud allows many possible shapes for the physical object boundary and it is hopeless, without additional assumptions, to infer the  usual pointwise curvature quantities on the physical object.

Indeed, a first difficulty is that the geometrical and topological properties of a physical object have to be considered at some  scale: for example, if one is interested in the shape of a ship hull, it may make sense to see it as a smooth surface at a large scale ($10^{-1}$ meters). However, at a finer scale
($10^{-4}$ meters) this same object may appear with many sharp features  near the rivets and small gaps between assembled sheets of metals.  At a still finer scale, one could even consider each atom as a separate connected component of the same physical object. 

A second difficulty is related to the ``pointwise" character of usual curvature definition. If the physical object is known up to $\epsilon$ in Hausdorff distance, it seems impossible to distinguish between the pointwise curvature at two points whose distance is of the order of $\epsilon$.
Without very strong assumptions on the regularity of the unknown object,  it is again meaningless to evaluate a pointwise curvature quantity.

In this work, we overcome the first difficulty by considering a scale dependent notion of curvature which consists merely, for a compact set $K$, in looking at the curvature of the $r$-offsets of $K$ ({\it i.e.} the set of points at distance less than or equal to $r$ from $K$). This offset, which can be seen as a kind of convolution, in the same spirit of similar operators in mathematical morphology or image processing, filters out high frequencies features of the object 's boundary. 

We overcome the second difficulty by considering curvature measures instead of pointwise curvature. In the case of smooths manifolds, curvature measures associate to a subset of the manifold the integral of pointwise curvatures over the subset.
However the curvature measures are still defined on non smooth objects such as convex sets  or more generally sets with positive reach \cite{federer-59}. 

In practice, in order to state our stability theorem, one has to make some assumption on the unknown physical object, namely the {\em positive $\mu$-reach} property defined below. To be more precise, our stability theorem still applies to objects whose offsets have positive $\mu$-reaches.

Some other approaches assume stronger properties such as smoothness or positive reach. But is it legitimate to make such assumptions on an unknown object? In practice, the only available informations about the physical object appears through the physical measures.
A distinctive character of our assumption on the $\mu$-reach  of offsets of physical object is that, thanks to the so-called critical values separation theorem \cite{CCSL06}, it can be reliably checked from the measured point sample. In this situation our curvature estimations reflect reliably the intrinsic properties of the physical object.

\paragraph{Related previous works.}
Due to its applications in geometry processing, many methods have been suggested that, given a triangulated surface, are able to estimate the curvature of an assumed underlying smooth surface (see \cite{PetitJean2002} for a survey).  Several authors (for example \cite{cazalspouget2003}) compute the curvature of a smooth polynomial surface approximating locally the triangulated surface. Closely related to our work is \cite{morvan-generalized,cohen-steiner,cohen-steiner-morvan-socg03,cohen-steiner-morvan-differentialgeometry}, which study a general definition of curvature measure that applies to both smooth surfaces and their approximation by triangulated surfaces, based on the so-called normal cycle (defined below). The proximity of curvature measure is proved using the powerful notion of {\em flat norm} between the corresponding normal cycles. These ideas are thoroughly reused in the present work. 

In \cite{CCSL06}, in order to address the question of topology determination through Hausdorff approximation, the authors  have introduced the class of sets with positive $\mu$-reach, which can be regarded as a mild regularity condition.
In particular, this condition does not require smoothness. If a set has a positive $\mu$-reach, or at least if it has offsets with positive $\mu$-reach, it is possible to retrieve the topology of its offset from the topology of some offsets of a Hausdorff close point sample.  A stable notion of normal cone has also be defined on this class of sets \cite{NormalConeApproximation}.
More recently, the authors have proved \cite{doubleOffset}  that the complement of offsets of sets with positive $\mu$-reach have positive reach. Since the normal cycle and the associated curvature measures are defined for sets with positive reach \cite{federer-59,fu-89} we may consider normal cycles of offsets of sets with  positive $\mu$-reach.
This paper develops this idea and gives a stability result whose proof uses the fact that the boundary of the double offsets of sets with positive $\mu$-reach are smooth surfaces.

\paragraph{Contributions.}
Our main result states that if the Hausdorff distance between two compact sets with positive $\mu$-reach is less than $\epsilon$, then the curvature measures of their offsets  differs by less that $O(\sqrt \epsilon)$, using an appropriate notion of distance between measures. This is then extended through the {\em critical values separation theorem} to the case where only one set has positive $\mu$-reach, which allows to evaluate the curvature measure of an object from a noisy point cloud sample. These results improve on the stability results in \cite{morvan-generalized,cohen-steiner,cohen-steiner-morvan-socg03,cohen-steiner-morvan-differentialgeometry}, which were only limited to the approximation of smooth hypersurfaces by homeomorphic triangulated manifolds. In order to provide a concrete algorithm we give the formulas that express these curvatures measures on a union of balls ({\it i.e.} on an offset of the point cloud). Closest to our work is \cite{boundarymeasure}, which also gives a stability result for curvature measures. The main differences are that our result also applies to anisotropic curvature measures, whereas \cite{boundarymeasure} is only limited to the usual curvature measures. On the other hand, the stability result for curvature measures in \cite{boundarymeasure} derives from a stability result for so-called boundary measures, which holds without any assumptions on the underlying compact set, whereas ours requires to assume a lower bound on the $\mu$-reach. While the two results seem related at first sight, the proof techniques are drastically different.

\paragraph{Outline.}
The paper first recalls definitions and properties related to the distance function to a compact set and its gradient, the critical function and the $\mu$-reach. Then, in Section \ref{section-currents}, one recalls classical notions concerning currents and one introduces the notions of normal cycle and of curvature measures. In Section \ref{section-main}, we state the stability theorem for sets with positive $\mu$-reaches. We then extend this theorem to the case where only one set has positive $\mu$-reach. In Section \ref{section-application},  we give the expressions of curvature measures for unions of balls. 
The last section gives the main steps of the proof of the main Theorem.

\section{Definitions and background on distance functions}\label{section-background}
We are using the following notations in the sequel of this paper.
Given $X \subset \R^d$, one denotes by $X^c$ the complement of $X$, by $\overline{X}$ its closure and
by $\partial X$ the boundary of $X$.
Given $A \subset \R^d$, $ch(A)$ denotes the convex hull of $A$.
\vskip5pt\noindent

\noindent
The \emph{distance function} $d_K$ of a compact subset $K$ of $\R^d$
associates to each point $x\in \R^d$ its distance to $K$:$$x\mapsto
d_K(x)= \min_{y \in K} d(x,y),$$ where $d(x,y)$ denotes the euclidean
distance between $x$ and $y$.  Conversely, this function characterizes
completely the compact set $K$ since $K=\{x\in \R^d \,|\, d_K(x)=0\}$.
Note that $d_K$ is $1$-Lipschitz. For a positive number $r$, we
denote by $K_r$ the $r$-offset of $K$, defined by $K_r
= \{x\,|\, d_K(x) \leq r \}$. The \emph{Hausdorff distance} $d_H(K, K')$ between two
compact sets $K$ and $K'$ in $\R^d$ is the minimum number $r$ such
that $K \subset K_r'$ and $K' \subset K_r$. It is not difficult to
check that the Hausdorff distance between two compact sets is the
maximum difference between the distance functions associated with the
compact sets: $$ d_H(K,K') = \sup_{x\in \R ^n} |d_K (x) - d_{K'} (x)|
$$

\noindent
Given a compact subset $K$ of $\R^d$, the {\em medial axis} $\mathcal{M}(K)$ of  $K$
is the set of points in $\R^d \setminus K$ that have at least two closest points on $K$. 
The infimum distance between $K$ and $\mathcal{M}(K)$ is called,
according to Federer, 
the {\em reach} of $K$ and is denoted $\reach(K)$. $\reach(K)=0$ if $K$ has concave sharp edges or corners.
The projection map $p_K$ that associates to a point $x$ its closest point $p_K(x)$ on $K$ is thus defined on $\R^d \setminus \mathcal{M}(K)$.

A $\C^{1,1}$ function is a $\C^{1}$ function
whose first derivative is Lipschitz.
 A $\C^{1,1}$ hypersurface $S$ is a $(d-1)$-manifold embedded in $\R^d$ such that
 each point of $S$ has a neighborhood which is the regular 
image (that is the image by a function whose derivative has  maximal rank) by an injective 
$\C^{1,1}$ function of a neighborhood of $0$ in $\R^{d-1}$. Informally,
one can say that a $\C^{1,1}$ surface is a surface with bounded curvature, which is strictly
stronger than $C^1$ and strictly weaker than $C^2$.
An embedded  $\C^{1}$
compact manifold is $\C^{1,1}$ if and only it has positive reach (Federer).

\subsection{The gradient} \label{section:GradientAndFlow}
\begin{figure}[!h]
\centerline{\includegraphics[height=6.5cm]{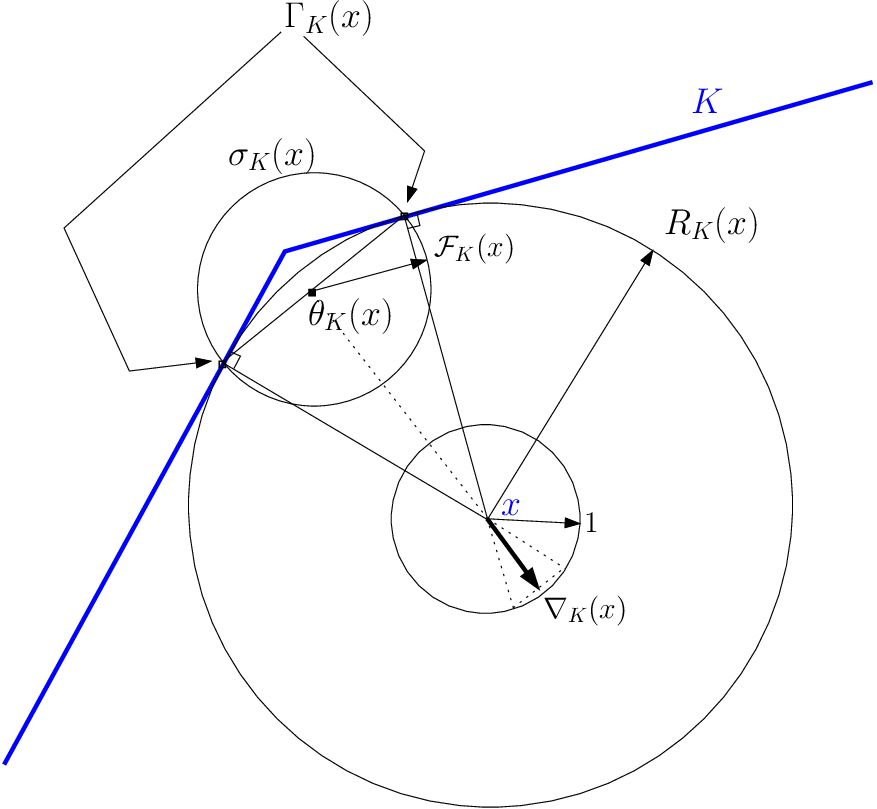}}
\caption{A 2-dimensional example with 2 closest points.} \label{fig:defNabla}
\end{figure}

The distance function $d_K$ is not differentiable on $\mathcal{M}(K)$. 
However, it is possible \cite{Li} to define a {\em generalized gradient} function $\nabla_K : \R^d \rightarrow \R^d$ that coincides with the usual gradient of $d_K$ at points where $d_K$ is differentiable.
For any point $x \in \R^d \setminus K$, we denote by  $\NearestSet_K(x)$ the set of points  in $K$ closest to $x$ (Figure \ref{fig:defNabla}):
\begin{equation*}
\NearestSet_K(x) = \{y \in K \,|\, d(x,y) = d_K(x) \}
\end{equation*}
Note that  $\NearestSet_K(x)$ is a non empty compact set. 
There is a unique smallest closed ball $\sigma_K(x)$ enclosing $\NearestSet_K(x)$  (cf. Figure \ref{fig:defNabla}).
We denote by $\theta_K(x)$  the center of  $\sigma_K(x)$ and by $\mathcal{F}_K(x)$ 
its radius.
$\theta_K(x)$  can equivalently be defined as the point on the convex hull of
$\NearestSet_K(x)$ nearest  to $x$.
For $x\in \R^d \setminus K$,  the generalized gradient $\nabla_K(x)$ is defined as follows:
\[
\nabla_K (x) = \frac{x-\theta_K(x)}{d_K(x)}
\]
It is natural to set  $\nabla_K (x)=0$  for $x \in K$.
Note that for $x\in \R^d\setminus K$, $||\nabla_K (x)||$ is the cosine of the (half) angle of the smallest cone with apex $x$ that contains $\NearestSet_K(x)$. 
 
\subsection{Critical points, critical function and the $\mu$-reach}
The {\em critical points} of $d_K$ are defined as
 the points $x$ for which $\nabla_K(x)=0$.
Equivalently,
a point $x$ is a  critical point if and only if it lies in the
convex hull of 
$\NearestSet_K(x)$. When $K$ is finite, this last definition means that critical points
are  precisely the intersections of Delaunay $k$-dimensional simplices
 with their dual $(d-k)$-dimensional Voronoi facets \cite{Giesen-John}.
Note that this notion of critical point is the same as the one considered in
 the setting of non smooth analysis \cite{CLARKE1} and Riemannian geometry \cite{CHEEGER,grove}.

The  results of this paper rely strongly
 on the notions of {\em critical function} and {\em $\mu$-reach},
introduced in \cite {CCSL06}. 
\begin{definition}[critical function]
Given a compact set $K\subset \R^d$, its \emph{critical function} $\chi_K:(0,+\infty) \to \R_+$ is the real function defined by:
$$\chi_K(d) = \inf_{d_K^{-1}(d)} ||\nabla_K||$$
\end{definition} 
The function $\chi_K$ is lower semicontinuous. The $\mu$-reach of a compact set $K$ is the maximal offset value $d$ for which $\chi_K(d')\geq \mu$ for $d' < d$.
\begin{definition}[$\mu$-reach]
The \emph{$\mu$-reach} $\rr_{\mu}(K)$ of a compact set $K\subset \R^d$ is defined by:
$$\rr_{\mu}(K) = \inf \{d \;|\;\chi_K(d) < \mu\}$$
\end{definition} 
We have that $\rr_{1}(K)$ coincides with the reach introduced by
Federer \cite{federer-59}.
 The critical function is in some sense ``stable'' with respect to small (measured by Hausdorff distance)
perturbations of a  compact set \cite {CCSL06}. That implies the following theorem \cite {CCSL06}:
\begin{theorem}[critical values separation theorem]\label{theorem:criticalValuesSeparation}
Let $K$ and $K'$ be two compact subsets of $\R^d$, $d_H(K,K') \leq \e$ and $\mu$ be a non-negative number. The distance function $d_K$ has no critical values in the interval $\left]4\e/\mu^2,r_{\mu}(K')-3\e\right[$. Besides, for any $\mu'<\mu$, $\chi_K$ is larger than $\mu'$ on the interval
$$
\left]\frac{4\e}{(\mu-\mu')^2},r_{\mu}(K')-3\sqrt{\e r_{\mu}(K')}\right[.
$$
\end{theorem}

\subsection{Complement of offsets and double-offset}
It has been proved in \cite{doubleOffset} that the complement $\overline{K_r^c}$ of the offset $K_r$ has positive reach
for any value $0 < r < r_\mu$. Moreover, one has a lower bound for
the critical function of $\overline{K_r^c}$:

\begin{theorem} \label{thm:offsetCriticalFunction}
For $r \in (0,r_\mu)$, one has 
\begin{equation} \label{eq:reach}
\reach (\overline{K_r^c}) \geq \mu r
\end{equation}
Moreover for any $t \in (\mu r, r)$,
\begin{equation} \label{eq:criticalfunction}
\chi_{\overline{K_r^c}}(t) \geq \frac{2\mu r - t(1+\mu^2)}{t(1-\mu^2)}.
\end{equation}
\end{theorem}
Let $K \subset \R^d$ be a compact set. For $0 < t < r$, 
the $(r,t)$-double offset $K_{r,t}$ of $K$ is the set defined by:
$$
K_{r,t}= \overline{(K_r)^c}_{\, t}.
$$
Using a result of Federer (\cite{federer-59}, Theorem 4.8) stating that the distance function to  a closed set $A$ with positive reach $r_1$ is differentiable with
non-zero Lipschitz gradient on the complement of the closure of the medial axis
of $A$,  the following result is obtained in \cite{doubleOffset}:

\begin{theorem} [Double offset theorem]\label{theorem:DoubleOffsetIsSmooth}
If $r < r_\mu$ for some value $\mu >0$ and if $t < \mu r $ then 
$\partial K_{r,t}$ is a smooth $\C^{1,1}$-hypersurface.
Moreover, 
$$
\reach(\partial K_{r,t}) \geq \min(t, \mu r -t)
$$
which implies that the smallest of the principal radii of curvature at
any point of $\partial K_{r,t}$ is at least $\min(t, \mu r -t)$.
\end{theorem}

\section{Definitions and background on curvature measures}\label{section-currents}
For more details on that section, one may refer to \cite{cohen-steiner,federer, morvan-generalized}. All the notions presented here are used in the proof of the main result (Theorem \ref{thm:main}). Theorem  \ref{thm:main} gives a result of stability for the curvature measures. The reader only interested in the result stated in Theorem \ref{thm:main} can skip most of this section and go directly to the subsection \ref{subsection-deuxiemeDefinition}: in that section, we give another way of defining the curvature measure that does not use the notion of currents and of normal cycles.

\subsection{General currents}
Let ${\mathcal D}^{m}$ the $\R$-vector space of  $C^\infty$ {\it 
differential $m$-forms} with compact support on $\R^n$ (see \cite{spivak} for details on differential $m$-forms). 
 ${\mathcal D}^{m}$ can be endowed with a topology similar to the topology on
the space of test functions used to define distributions, the
so-called $C^\infty$ topology. A sequence $(\phi_i)$ of elements of ${\mathcal
D}^{m}$ converges to $\phi \in {\mathcal D}^{m}$ in the $C^\infty$
topology if and only if there exists a compact $K$ containing the supports of all the $\phi_i$ such that the derivatives of any order of the $\phi_i$ converge to the corresponding derivatives of $\phi$ uniformly in $K$. \\
The topological 
dual of ${\mathcal D}^{m}$ is the $\R$-vector space ${\mathcal D}_{m}$ of 
{\it $m$-currents} on $\R^n$. Equivalently, currents can be viewed as 
differential forms
whose coefficients are distributions instead of smooth functions.
The {\it support} $Spt(T)$ of a current $T$ can be defined as the
union of the support of its coefficients.

The subset of $m$-currents 
with compact support is denoted by ${\mathcal E}^{m}$. We endow ${\mathcal
D}_{m}$ with the weak topology: 
 
 \begin{equation*}\label{N3} 
 \lim_{j \to \infty} T_{j}= T \iff 
 \forall \phi \in {\mathcal D}^{m} \lim_{j \to \infty}T_{j}(\phi)= T(\phi) 
 \end{equation*} 

\subsubsection{Operations on currents}  
Basic notions relative to differential forms can be transposed to
currents by dualization~:
\begin{enumerate} 
\item{Boundary : } to each $m$-current $T$ one can associate a
$m-1$-current $\partial T$ called its {\it boundary}, defined by~:
     $$\forall \phi \in {\mathcal D}^{m-1}\; \partial T(\phi) = T(d\phi)$$
where $d\phi$ denotes the exterior derivative of $\phi$ (see \cite{spivak} for the definition).
\item{Push-forward : } given an $m$-current $T$ and a smooth map $f$
defined on a neighborhood of the support of $T$, one can define the 
{\it push-forward} $f_\sharp T$ of $T$ by $f$~:
$$\forall \phi \in {\mathcal D}^{m}\; f_\sharp T(\phi) = T(f^* \phi)$$
where $f^* \phi$ is the pull-back of $\phi$ by $f$ (see \cite{spivak}). Note that this 
definition only makes sense when $f^* \phi$ is compactly
supported. We thus have to assume that $f$ is proper, that is
$f^{-1}(K)$ is compact for every compact $K$. Actually, for the
push-forward to be defined, it is sufficient that the restriction of
$f$ to the support of $T$ is proper.
\end{enumerate} 
Since $f^*$ commutes with exterior differentiation, $f_\sharp$
commutes with the boundary operator, so that the push-forward of a
current without boundary is also without boundary.

\subsubsection{Current representable by integration} 
We say that a current $T \in {\mathcal D}_{m}$ is 
{\it representable by integration} if there is a Borel 
regular measure $||T||$ on $\R^{n}$ finite on compact 
subsets and a unit $m$-vector fields $\overrightarrow{T}$ 
defined almost everywhere such that   
 
\begin{equation*}\label{PM} 
\forall \phi \in {\mathcal D}^{m}\; T(\phi) = \int<\overrightarrow{T},\phi
>d||T||  
\end{equation*} 

Currents representable by integration are analogous to distributions of 
order $0$. A current representable by integration $T$ can be ``restricted''
to any $||T||$-measurable set $A$ (see \cite{federer} pp 356). The obtained 
current $T\llcorner A$ is defined by~:
\begin{equation*}\label{PMR} 
\forall \phi \in {\mathcal D}^{m}\; T{\llcorner A}(\phi) = \int<\overrightarrow{T},\phi
>{\bf 1}_A\;d||T||  
\end{equation*} 
 
    \subsubsection{Rectifiable and integral currents} 
In particular, one can associate an $m$-current representable by
integration to any oriented $m$-rectifiable subset $S$ of dimension $m$ of 
$\R^n$ (see \cite{federer}). It is a well-known fact that rectifiable sets of dimension $m$ 
have a well-defined tangent space at ${\mathcal H}^m$-almost every point. Let $\vec{S}$ be 
the unit $m$-vector field encoding these -oriented- tangent spaces. 
The current associated with $S$, still 
denoted by $S$, is defined by~: 
 \begin{equation*}\label{S1} 
S(\phi)=\int_{S} <\vec{S},\phi>d{\mathcal H}^m 
 \end{equation*} 
More general currents can be defined by incorporating integer
multiplicities $\mu$ in the previous formula~:
 \begin{equation*}\label{S2} 
T(\phi)=\int_{S} \mu <\vec{S},\phi>d{\mathcal H}^m 
 \end{equation*} 
If the support of $S$ is compact, and $\int_S\mu d{\mathcal H}^m<\infty$,
we say that $T$ is {\it rectifiable}. The space of rectifiable 
currents is denoted by ${\mathcal R}_{m}$.\\ 
 
A current is said to be {\it integral} if it is rectifiable and if 
its boundary is rectifiable. 
  
\subsubsection{Mass and norms of currents} 
The norm of 
a $m$-differential form $\phi$ is the real number 
 
 \begin{equation*}\label{NN1} 
 ||\phi||= \sup_{p \in M^{n}} ||\phi_{p}||, 
 \end{equation*} 
 where, for each $p\in M^{n}$, 
 
 \begin{equation*}\label{N2} 
 ||\phi_{p}||= \sup \{|<\phi_{p}, \zeta_{p}>|, \zeta_{p} \in 
 \Lambda^{m}T_{p}M^{n}, |\zeta_{p}|=1\}. 
  \end{equation*} 
There are different interesting norms on the  space of 
currents ${\mathcal D}_{m}$. We mention the main ones: 
    \begin{itemize} 
        \item 
            The mass of a current $T\in {\mathcal D}_{m}$ is the real 
            number 
             \begin{equation*}\label{MU} 
            {\bf M}(T)= \sup\{T(\phi), \mbox{ such that } \phi \in {\mathcal D}^{m}, ||\phi||\leq 
            1.\} 
             \end{equation*} 
For rectifiable currents of dimension $m$, the mass somehow
generalizes the notion of $m$-volume~: the mass of the current $T$ defined by
\ref{S2} is $\int_S\mu d{\mathcal H}^m$. Rectifiable currents thus have
finite mass. 
Using general results on representation theory of geometric measure theory, 
it can be proved that if ${\bf M}(T)<\infty$,  $T$ is representable by integration. 
        \item 
            The flat norm of a current $T\in {\mathcal D}_{m}$ is the real 
            number 
             \begin{equation*}\label{FF} 
             {\mathcal F}(T)= \inf \{{\bf M}(A) + {\bf M}(B) \mbox{ such that } T= A + \partial B , A\in {\mathcal R}_{m}, 
             B\in {\mathcal R}_{m+1}\}. 
              \end{equation*} 
    \end{itemize} 
 
It can be shown that the flat norm can also be expressed in the
following way~: 
    
              \begin{equation*}\label{FH} 
{\mathcal F}(T)= \sup\{T(\phi), \mbox{ such that } \phi \in {\mathcal 
D}^{m}, ||\phi||\leq 
            1, ||d\phi||\leq 
            1\}. \end{equation*}

\subsection{Normal cycle of geometric sets}
For many compact sets of $\R^d$, one can associate a (d-1)-current that generalizes the notion of unit normal bundle of a smooth manifold (see \cite{fu-89} for more details). In the case  where $\C$ is a compact set of $\R^d$ enclosed by a smooth (d-1)-dimensional manifold,  the normal cycle $N(\C)$ of $\C$ is just the integral (d-1)-current associated to its outer unit normal bundle:
$$
S(\C)=\{(p,n(p)),\ p\in \partial \C,\ n(p) \mbox{ is the outer unit normal at }p\}.
$$
If $\C$ is a convex set, the normal cycle $N(\C)$ is just the integral (d-1)-current associated to the oriented set:
$$
S(\C)=\{(p,n),\ p\in \partial \C,\ n \in CN_{\C}(p)\},
$$
where $CN_{\C}(p)=\{n\in \S^2,\ \forall x \in \C\ n.\overrightarrow{px}\leq 0\}$ is the normal cone of $\C$ at the point $p$.
If $\C=\cup_{i=1}^n \C_i$  is a union of convex sets,  its normal cycle is defined by additivity. By using the inclusion-exclusion formulae, we have:
$$
N(\C)=\sum_{l=1}^n(-1)^{l+1}\sum_{1\leq i_1 < ...<i_l \leq n}N(\cap_{j=1}^l\ds \C_{i_j}).
$$
The last definition implies that the normal cycle is well defined for polyedron, but also for a finite union of balls. In fact, Joseph Fu \cite{fu-89} proved that the normal cycle can be generalized to a very broad class of objects, which he calls {\it geometric sets}. This class of objects contains in particular subanalytic sets \cite{fu-54}, definable sets \cite{bernig-brocker}, riemannian polyedra \cite{hug-schneider}, or sets with positive reach \cite{fu-89}.
\subsection{Curvature measures}
Let us first recall some basic definitions and notations in the case where $M$ is a smooth surface that is the boundary of a compact set $V$ of $\R^d$. The unit normal vector at a point $p\in M$ pointing outward $V$ will be refered as $n(p)$. Note that $M$ is thereby oriented. Given a vector $v$ in the tangent space $T_pM$ to $M$ at $p$, the derivative of $n$ in the direction $v$ at $p$ is orthogonal to $n(p)$. The derivative $D_pn$ of $n$ at $p$ thus defines an endomorphism of $T_pM$, known as the Weingarten endomorphism. The Weingarten endomorphism is symmetric. The associated quadratic form is called the second fundamental form. Eigenvectors and eigenvalues of the Weingarten endomorphism are respectively called principal directions and principal curvatures. In the $3$-dimensional case, both principal curvatures can be recovered from the trace and determinant of $D_pn$, also called mean and gaussian curvature.

The curvature measures can be defined for any compact of $\R^d$ admitting a normal cycle ({\it i.e.} for geometric sets). Before introducing their definition, we first need to give the definition of invariant forms as follows.

We identify the tangent bundle $T\R^d$ with $E\times F$, where $E$ is the base space and $F$ is the fiber. Let $J:E\to F$ be the canonical isomorphism  between $E$ and $F$. We endow $T\R^d$ with the dot product $<(e,f),(e',f')>=<e,e'>+<J^{-1}(f),J^{-1}(f')>$. At any point $(m,\xi)$ of $ST\R^d=\{(m,\xi)\in E\times F,\ \|\xi\|=1\}$, we consider an orthonormal frame $(e_1,...,e_{d-1})$ of the space orthogonal to $J^{-1}(\xi)$ and we take $\epsilon_i=J(e_i)$. We build the (d-1)-differential form: 
$$
\Omega=(e_1^*+t\epsilon_1^*)\wedge...\wedge(e_{d-1}^*+t\epsilon_{d-1}^*),
$$
where $u^*$ denotes the 1-form defined by $u^*(x)=<u,x>$. One can show that this form does not depend on the chosen orthonormal frame. The coefficient of $t^i$ is a (d-1)-form  denoted by $\omega_i$. One can show that each $\omega_i$ is invariant under the action of the orthogonal group.
We now define the curvature measures as follows:
\begin{definition}
Let $\C$ be a geometric compact subset of $\R^d$. The $k^{th}$-curvature measure of $\C$, denoted by $\Phi^k_{\C}$ associates to each Borel subset $B$ of $\R^d$ the real number:
$$
\Phi^k_{\C}(B)=N(\C)\llcorner (B\times F) (\omega_k).
$$
\end{definition}
Note that $\Phi^k_{\C}(B)=N(\C)(\ind_{B\times F}\ \omega_k)$, where $\ind_{B\times F}$ is the indicatrix function of $B\times F$. Therefore, we can extend this notion of curvature measure to any Lipschitz real function. This point of view is crucial and is a key point that will allow us to state simple results of stability in this paper. More precisely, one defines the isotropic  curvature measure for every Lipschitz function $f$ on $\R^d$ by:
$$
\Phi^k_{\C}(f)=N(\C) (\bar{f}\ \omega_k),
$$
where $\bar{f}$ is defined on $\R^d\times \R^d$ 
by ${\bar{f}}(p,n)=f(p)$.  

One can show that if $\C$ is the volume enclosed by an hypersurface $\partial \C$, then $\Phi^k_{\C}(B)$ is the integral over $\partial \C \cap B$ of the k-th symmetric function  of the principal curvatures of $\partial \C$ \cite{morvan-generalized}. 

In dimension $3$, $\Phi^1_{\C}$ and $\Phi^2_{\C}$ are respectively the integral of twice the mean curvature and the integral of the gaussian curvature. We then take the notation $\Phi^G_{\C}=\Phi^2_{\C}$ and $\Phi^H_{\C}=\Phi^1_{\C}$. One has:
$$
\Phi^G_{\C}(B)=\int_{B\cap \partial \C} G(p)dp
\quad and \quad
\Phi^H_{\C}(B)=\int_{B\cap \partial \C} H(p)dp,
$$
where $H(p)$ an $G(p)$ denote respectively the gaussian and the mean curvature of $\partial \C$ at $p$.

\subsection{Anisotropic curvature measure}
The anisotropic curvature measure \cite{cohen-steiner, cohen-steiner-morvan-differentialgeometry,morvan-generalized} is defined as follows: 
\begin{definition}
Let $\C$ be a geometric subset of $\R^d$. The anisotropic curvature measure of $\C$, denoted by $\overline{H}_{\C}$ associates with every Borel subset $B$ of $\R^3$ the following bilinear form on $E$:
$$
\forall X,Y \in \R^d\ \overline{H}_{\C}(B)(X,Y)=N(\C)\llcorner(B\times F)(\omega^{X,Y}),
$$
where
$$
\omega^{X,Y}=*_E(J^{-1}(\xi)^* \wedge X^*)\wedge J(Y)^*,
$$
here $*_E$ denotes the Hodge dual on $E$.
\end{definition}
In the particular case where $\C$ is the volume enclosed by a $C^2$ compact hypersurface,  $\overline{H}_{\C}(B)$ is just the integral over $B\cap \partial \C$ of a symmetric bilinear form $H_{\C}$ related to the second fundamental form of $\partial \C$. More precisely, this form $H_{\C}$ coincides with the second fundamental form of $\partial \C$ on the tangent plane of $\C$ and vanishes on its orthogonal complement. For any Borel set $B$ of $\R^d$, one has \cite{cohen-steiner,cohen-steiner-morvan-differentialgeometry,morvan-generalized}:
$$
\overline{H}_{\C}(B)=\int_{B\cap\partial \C}H_{\C}(p)dp.
$$

In the $3$-dimensional case, it is also convenient to introduce another form $\widetilde{\omega}^{X,Y}$ useful, in particular for polyedrons \cite{cohen-steiner-morvan-socg03}. At a given point $(m,\xi)\in\R^3\times\S^2$, one has: 
$$
\omega^{X,Y}_{(m,\xi)}=(0,Y)^*\ \wedge (\xi\times X,0)^*
\quad \mbox{and} \quad
\widetilde{\omega}^{X,Y}_{(m,\xi)}=(X,0)^*\ \wedge (0,\xi\times Y)^*,
$$
where $\times$ is the cross-product in $\R^3$. It is then possible to define another anisotropic curvature measure, denoted by $\overline{\widetilde{H}}_{\C}$, that associates with every Borel subset $B$ of $\R^3$ the following bilinear form on $E$:
$$
\forall X,Y \in \R^d\ \overline{\widetilde{H}}_{\C}(B)(X,Y)=N(\C)\llcorner(B\times F)(\omega^{X,Y}).
$$
In particular, if $\C$ is the volume enclosed by a $C^2$ compact surface, one has:
$$
\overline{\widetilde{H}}_{\C}(B)=\int_{B\cap\partial \C}\widetilde{H}_{\C}(p)dp,
$$
where $\widetilde{H}_{\C}$ is defined as having the same eigenvectors than ${H}_{\C}$, but with swapped eigenvalues on the tangent plane, and vanishes on the orthogonal component of $\C$.

Similarly as in the previous subsection, one can also define the anisotropic  curvature measures $\overline{{H}}_{\C}(f)$ and $\overline{\widetilde{H}}_{\C}(f)$ for every Lipschitz function $f$ on $\R^d$.

\subsection{Another approach for defining curvature measures}\label{subsection-deuxiemeDefinition}
It is possible to define the curvature measures of a set with positive reach as the limit of the curvature measures of its offsets. The advantage of these definitions is that they do not rely on the notion of normal cycle.

More precisely, let $V$ be a set with positive reach $R>0$ and let $t<R$. It is known that $\partial V_t$ is a $\C^{1,1}$ hypersurface of $\R^d$ \cite{federer-59}.  The second fundamental form and the principal curvatures of $\partial V_t$ are thus defined almost everywhere. There is of course no pointwise convergence of the principal curvatures when $t$ tends to $0$. However, the integrals of the curvatures of $\partial V_t$ converge to the integrals of the curvatures of $V$ when $t$ tends to $0$ (this is a consequence of  Lemma \ref{lemma-erosions}). That allows us to define the isotropic curvature measures of $V$ for every Borel subset $B$ of $\R^d$ as follows:
$$
\Phi^k_{V}(B)=\lim_{t\to 0} \int_{\partial V_t\cap B'}s^k(p)dp
$$
where $p_V$ is the projection onto $V$; $B'=\{p\in \R^d,\ p_V(p)\in B\}$; $s^k$ is the k-th elementary symmetric polynomial of the principal curvatures $\lambda_1$,...,$\lambda_{d-1}$ of $\partial V_t$. In other words, $s^k$ satisfies for every $x\in \R$:
$
(x+\lambda_1)...(x+\lambda_{d-1})=s^0+s^1x+...+s^{d-1}x^{d-1}.
$
Now, remark that we have:
$$
\Phi^k_{V}(B)=\lim_{t\to 0} \int_{\partial V_t}\ind_B(p_V(p))\ s^k(p)dp,
$$ 
where $\ind_B$ is the indicatrix function of $B$. Therefore, we can extend this notion of curvature measure to any Lipschitz real function. This point of view is crucial and is a key point that will allow us to state simple results of stability in this paper. More precisely, one defines the isotropic  curvature measure for every Lipschitz function $f$ on $\R^d$ by:
$$
\Phi^k_{V}(f)
=\lim_{t\to 0} \int_{\partial V_t}f(p_V(p))\ s^k(p)dp.
$$
Similarly, one extends the notion of anisotropic curvature measure of \cite{cohen-steiner,cohen-steiner-morvan-socg03}: the anisotropic curvature measure of $V$ associates to any Lipschitz function $f$ the $d\times d$ symmetric matrix defined by:
$$
\overline{H}_{V}(f)=\lim_{t\to 0} \int_{\partial V_t}f(p_V(p))\ H_{\partial V_t}(p)dp,
$$
where $H_{\partial V_t}$ is a matrix-valued function defined on $\R^d$ that coincides with the second fundamental form of $\partial V_t$ on the tangent space, and vanishes on the orthogonal component.

Now, let $K$ be a compact set whose $\mu$-reach is greater than $r>0$. Then $V=\overline{K_r^c}$ has a reach greater than $\mu r$. It is then possible to define the extended notions of curvature measures of $K_r$ by (see Lemma \ref{lemma-erosions}):
$$
\Phi^k_{K_r}(f)=(-1)^k\Phi^k_{V}(f) 
\quad \mbox{and} \quad 
 \overline{H}_{K_r}(f)= - \overline{H}_{V}(f).
$$
In the $3$-dimensional case, it is also possible to define a second anisotropic measure curvature $\overline{\widetilde{H}}$ by:
$$
\overline{\widetilde{H}}_{V}(f)=\lim_{t\to 0} \int_{\partial V_t}f(p_V(p))\ \widetilde{H}_{\partial V_t}(B)
\quad \mbox{and} \quad
 \overline{\widetilde{H}}_{K_r}(f)= - \overline{\widetilde{H}}_{V}(f),
$$
where $\widetilde{H}_{\partial V_t}$ is defined as having the same eigenvectors than ${H}_{\partial V_t}$, but with swapped eigenvalues on the tangent plane, and vanishes on the orthogonal component of $\partial V_t$.

\section{Stability results}\label{section-main}

\subsection{Curvature measures of the offsets}
The main contribution of this paper is Theorem \ref{thm:main}. Thanks to the formulation of the curvature measures with Lipschitz functions, the statement is simple. This theorem states that if two compact sets $K$ and $K'$ with positive $\mu$-reaches are close in the Hausdorff sense, then the curvature measures of their offsets are close.
We recall that the covering number ${\cal N}(A,t)$ of a compact set $A$ is the minimal number of closed balls of radius $t$ needed to cover $A$.

\begin{theorem} \label{thm:main}
Let $K$ and $K'$ be two compact sets of $\R^d$ whose $\mu$-reaches are greater than $r$. We suppose that the Hausdorff distance $\epsilon=d_H(K,K')$  between $K$ and $K'$ is less than $\frac{r\mu\ (2-\sqrt{2})}{2}\min(\mu,\frac{1}{2})$. If $f:\R^d \to \R$ is a Lipschitz function satisfying $|f| \leq 1$, then:
$$
|\Phi^i_{K_r}(f)-\Phi^i_{K'_r}(f)| \leq k(r,\mu,d,f)\ \sup(Lip(f),1)\ \sqrt{\epsilon},
$$
and
$$
\| \overline{H}_{K_r}(f)-\overline{H}_{K'_r}(f)\| \leq k(r,\mu,d,f)\ \sup(Lip(f),1)\  \sqrt{\epsilon},
$$
where $k(r,\mu,d,f)$ only depends on $f$ through the covering number ${\cal N}( spt(f)_{O(\sqrt{\epsilon})},\mu r/2)$; $Lip(f)$ is the Lipschitz-constant of $f$; $spt(f)=\{x\in\R^d,\ f(x)\neq 0\}$.
\end{theorem}

\begin{figure}[h!]
\centerline{\includegraphics[height=5cm]{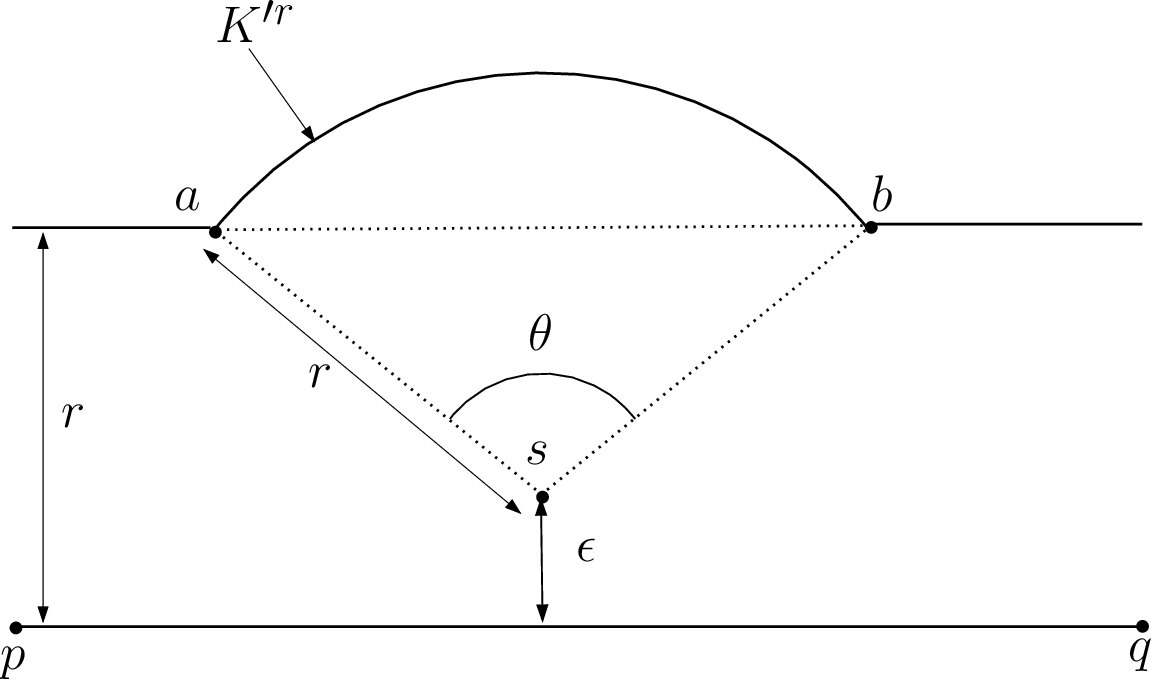}}
\caption{Tightness of the bound: we take $K=[p,q]$ and $K'=[p,q]\cup \{s\}$, where $s$ is at a distance $\epsilon$ from $K$. We have $d_H(K,K')=\epsilon$ and the total curvature $\theta$ of $K'_r$ between $a$ and $b$ satisfies
$\theta = 2 \arccos \left(\frac{r-\epsilon}{r}\right)=O\left(\sqrt{\epsilon}\right).$
}
\label{figure-tightexample}
\end{figure}

The proof of this theorem is given in Sections \ref{section-one-compact}, \ref{section-double-offsets} and \ref{section-double-offset-2-offset}. We show in Figure \ref{figure-tightexample} that this bound is tight. Furthermore, in the $3$-dimensional case, this result also holds for the anisotropic curvature measure $\overline{\widetilde{H}}$. 

Now, if we take the function $f(x)=\max(1-\|x-c\|/r,0)$ equal to $1$ at a point $c\in \partial K'_r$ that radially decreases in a ``small" ball $\BB$ of radius $r$ and vanishes out of $\BB$, then we can get local information about the curvature of $K'_r$ from the curvature of $K_r$ in the neighborhood of $c$.

We also note that the conclusion of the theorem may be rephrased by saying that the {\it bounded Lipschitz distance} between the curvature measures of $K_r$ and $K'_r$ is bounded by $O(\sqrt{\epsilon})$. The bounded Lipschitz distance between measures is similar to the Wasserstein distance (also called earth's mover distance), except that it applies to general signed measures whereas Wasserstein distance is limited to probability measures. We refer to \cite{boundarymeasure} for precise definitions. 

\subsection{General result}
The result of the previous  section ensuring the stability of the curvature measures assume that both the compact sets $K$ and $K'$ have sufficiently large $\mu$-reach. Nevertheless, in practical settings, particularly when dealing with point clouds, such an hypothesis is rarely satisfied. Using Theorem \ref{theorem:criticalValuesSeparation},  it is still possible to approximate the curvature measures of the offsets of a compact set with positive $\mu$-reach from any sufficiently close approximation of it.

\begin{theorem} \label{thm:general_result}
Let $K$ and $K'$ be two compact subsets of $\R^d$ such that $r_\mu(K') >r$. Assume that the Hausdorff distance $\e = d_H(K,K')$ between $K$ and $K'$ is such that $\e < \frac{\mu^2}{60 + 9 \mu^2} r$. 
Then the conclusions of Theorem \ref{thm:main} also hold.
\end{theorem}
Of course, if for $s>0$, the compact set $K'$ satisfies $r_{\mu}(K'_s)>c>0$, the same theorem applies to $K'_s$. Furthermore, thanks to Theorem \ref{theorem:criticalValuesSeparation}, the value of $r_{\mu}(K'_s)$ can be read on the critical function of the sample $K$.

\begin{proof}
It follows from Theorem \ref{theorem:criticalValuesSeparation} (the critical values separation theorem of \cite{CCSL06}) that the critical function of $K$ is greater than $\frac{\mu}{2}$ on the interval $\left ( \frac{16\e}{\mu^2}, r - 3 \sqrt{\e r} \right )$. Note that since $\mu > \frac{\mu}{2}$, the critical function of $K$ is also greater than $\frac{\mu}{2}$ on the same interval. As a consequence the two compact sets $\tilde K = K_{\frac{16\e}{\mu^2}}$ and $\tilde K' = K'_\frac{16\e}{\mu^2}$ have their $\frac{\mu}{2}$-reach greater than $r - 3 \sqrt{\e r} - \frac{16\e}{\mu^2}$. 
Notice that for any $\delta>0$, $\tilde K_\delta = K_{\frac{16\e}{\mu^2}+\delta}$ and since $d_H(K,K') = \e$ then $d_H(\tilde K, \tilde K') \leq \e$.
To apply Theorem \ref{thm:main} to $\tilde K$ and $\tilde K'$, $\frac{\mu}{2}$ and $r$, the Hausdorff distance $\e$ between $\tilde K$ and $\tilde K'$ must satisfy $0 < \e < \frac{2-\sqrt{2}}{8} \mu^2 (r - 3 \sqrt{\e r} - \frac{16\e}{\mu^2})$ (note that since $\frac{\mu}{2} \leq \frac{1}{2}$, $\min(\mu, \frac{1}{2}) = \mu$). One easily checks (by computing the solutions of the inequality and using that for all $0\leq x \leq 1$, $\sqrt{1-x} \geq 1 - \frac{x}{2}$) that this inequality is implied by the assumption made on $\e$. The theorem now follows immediately from Theorem \ref{thm:main}.
\end{proof}

Remark that in the previous proof, the choice of $\frac{\mu}{2}$ to apply the critical values separation theorem of \cite{CCSL06} and then the theorem \ref{thm:main} is arbitrary. The same proof can be given replacing $\frac{\mu}{2}$ by any $0 < \mu' < \mu$ leading to somewhat different, but more technical, constraints on $\e$ and $r$ in the statement of Theorem \ref{thm:main}.

\section{Computation of the curvature measures of $3D$ point clouds}\label{section-application}
\subsection{Gaussian and mean curvature measures of $3D$ point clouds}
Let $K$ be a finite set of points in $\R^3$ and $r>0$. We assume that the set of balls of radius $r$ and centered in $K$ are in general position (as in \cite{edelsbrunner-mucke}): no $4$ points lie on a common plane; no $5$ points lie on a common sphere; and the smallest sphere through any 2, 3 or 4 points of $K$ has a radius different from $r$. In the $3$-dimensional case, we denote by $\Phi^H_{\C}=\Phi^1_{\C}$ the mean curvature measure and by $\Phi^G_{\C}=\Phi^2_{\C}$ the gaussian curvature measure.

Note that the boundary of the union of balls $K_r$ is a spherical polyhedron: its faces are spherical polygons contained in the spheres of radius $r$ centered on points of $K$; its edges are arcs of circles contained in the intersection of pairs of spheres of radius $r$ centered on $K$; its vertices belong to the intersection of three spheres of radius $r$ and centered in $K$. It follows from Lemma 2.2 in \cite{edelsbrunner-1993} that the combinatorial structure of $\partial K_r$ is in one-to-one correspondence with the boundary of the $\alpha$-shape of $K$ for $\alpha = r$. As a consequence, under the general position assumption, to compute $\Phi^G_{K_r}(f)$ and $\Phi^H_{K_r}(f)$, it is sufficient to compute the curvature measures $\Phi^G_{\C}(f)$ and $\Phi^H_{\C}(f)$, where $\C$ is the union of one, two or three balls of radius $r$. The orientation of the boundary of the union of ball is taken so that the normal is pointing outside. 
All the curvature measures for the union of one, two or three balls are given respectively in Propositions \ref{proposition_1ball}, \ref{proposition-mesures-2-boules} and \ref{proposition-mesures-3-boules}.

\begin{proposition}\label{proposition_1ball}
Let $\BB$ be a ball of radius $r$ of $\R^3$ and let $B$ be a Borel set of $\R^3$. Then the curvature measures of $\BB$ above $B$ are given by:
$$
\Phi^H_{\BB}(B) = \frac{2}{r}\ Area(B\cap \partial \BB)
\quad and \quad
\Phi^G_{\BB}(B)
=\frac{1}{r^2}\ Area(B\cap \partial \BB).
$$
\end{proposition}
\begin{proof}
Since $\partial \BB$ is a smooth surface, one has
$$
\Phi^H_{\BB}(B) 
= \int_{B\cap \partial \BB} H(p)dp
= \frac{2}{r}\ Area(B\cap \partial \BB)
$$
and
$$
\Phi^G_{\BB}(B)
= \int_{B\cap \partial \BB} G(p)dp
=\frac{1}{r^2}\ Area(B\cap \partial \BB),
$$
where $G(p)$ is the Gaussian curvature and $H(p)$ is the mean curvature  of $\BB$ at $p$. 
\end{proof}
In the following, we use the notation: $\omega^H=\omega_1$ and $\omega^G=\omega_2$.
\begin{proposition}\label{proposition-mesures-2-boules}
Let $\BB_1$ and $\BB_2$ be two intersecting balls of $\R^3$, of same radius $r>0$ and   of centers $A_1$ and  $A_2$, and $C$ be the circle $\partial \BB_1 \cap \partial \BB_2$.  Let $B$ be a ball of $\R^3$. Then the curvature measures of $\BB_1\cup \BB_2$ above $B\cap C$ are given by:
$$
\Phi^H_{\BB_1\cup \BB_2}(B\cap C)
=-2\beta r\ arcsin\left(\frac{A_1A_2}{2r}\right) \sqrt{1- \left(\frac{A_1A_2}{2r}\right)^2}
$$
and
$$
\Phi^G_{\BB_1\cup \BB_2}(B\cap C)
=- \frac{\beta\ A_1A_2}{r},
$$
where $\beta=2\pi \frac{length(B\cap C)}{length(C)}$ is the angle of the arc of circle $B\cap C$.
\end{proposition}
\begin{proof}
Since the normal cycle  is additive and since $C$ is one-dimensional, one has:
$$
\begin{array}{rl}
\Phi^H_{\BB_1\cup \BB_2}(B\cap C)
&=\Phi^H_{\BB_1}(B\cap C)+\Phi^H_{\BB_2}(B\cap C)-\Phi^H_{\BB_1\cap \BB_2}(B\cap C)\\
&=-\Phi^H_{\BB_1\cap \BB_2}(B\cap C).
\end{array}
$$
We now need to describe the normal cycle of $\BB_1 \cap \BB_2$ ``above" $C\cap B$. Since $\BB_1\cap \BB_2$ is convex, its normal cycle ``above" $C\cap B$ is just the $2$-current defined by integration over the set:
$$
S_C(\BB_1\cap \BB_2)=\left\{(m,\xi),\ m\in C\cap B,\ \|\xi\|=1\ and\ \forall q \in \BB_1\cap \BB_2\ \overrightarrow{mq}.\xi \leq 0\right\}.
$$
Let $\alpha=arcsin\left(\frac{A_1A_2}{2r}\right)$. In an suitable frame, the set $S_C(\BB_1\cap \BB_2)$ can be parametrized by:
$$
\begin{array}{rlll}
f:&[0,\beta]\times[-\alpha,\alpha]&\to&\R^3\times \S^2\\
&(u,v)&\mapsto
&\left(\begin{array}{cc}
\left(\begin{array}{c}r \cos \alpha \cos u\\ r\cos \alpha \sin u\\ 0 \end{array}\right),
\sin(v)\ \left(\begin{array}{c}0 \\ 0\\ 1 \end{array}\right)
+\cos(v)\ \left(\begin{array}{c}\cos u\\ \sin u\\ 0 \end{array}\right)
\end{array}\right)
\end{array}
$$
Let $(m,\xi)=f(u,v) \in S(\BB_1\cap \BB_2)$. We put
$$
e_1=\left(\begin{array}{c} -\sin u\\ \cos u\\ 0 \end{array}\right) 
\mbox{ and }
e_2=\cos(v)\ \left(\begin{array}{c}0 \\ 0\\ 1 \end{array}\right)
-\sin(v)\ \left(\begin{array}{c}\cos u\\ \sin u\\ 0 \end{array}\right).
$$
Then $(e_1,e_2,\xi)$ is a direct orthonormal basis of $\R^3$. We put $\epsilon_1=(e_1,0)$, $\epsilon_2=(e_2,0)$, $\widetilde{\epsilon_1}=(0,e_1)$ and $\widetilde{\epsilon_2}=(0,e_2)$. On has \cite{cohen-steiner}:
$$
\omega^H=\epsilon_1 \wedge \widetilde{\epsilon_2} + \widetilde{\epsilon_1}\wedge \epsilon_2
\quad \mbox{and} \quad
\omega^G=\widetilde{\epsilon_1} \wedge \widetilde{\epsilon_2}.
$$
Furthermore, one has:
$$
\frac{\partial f}{\partial u}(u,v)=(r \cos \alpha\ e_1,\cos v\ e_1)
\quad \mbox{and} \quad
\frac{\partial f}{\partial v}(u,v)=(0,e_2).
$$
We then have:
$$
f^*\omega^H((1,0),(0,1))=\omega^H\left(\frac{\partial f}{\partial u}(u,v),\frac{\partial f}{\partial v}(u,v) \right)=r \cos \alpha,
$$
$$
f^*\omega^G((1,0),(0,1))=\cos v.
$$
Then
$$
\Phi^H_{\BB_1\cup \BB_2}(C)
=- \int_{0}^{\beta}\int_{-\alpha}^{\alpha}r\cos \alpha\ dudv
=-2\beta r \alpha \cos \alpha.
$$
Similarly, one has:
$$
\Phi^G_{\BB_1\cup \BB_2}(C)
=- \int_{0}^{\beta}\int_{-\alpha}^{\alpha}\cos v\ dudv
=- 2\beta\sin \alpha
=- \frac{\beta\ A_1A_2}{r}.
$$
\end{proof}
\begin{proposition}\label{proposition-mesures-3-boules}
Let $\BB_1$, $\BB_2$ and $\BB_3$ be three intersecting balls of $\R^3$, of radius $r$,  of centers $A_1$, $A_2$ and $A_3$, and $p\in \partial \BB_1\cap  \partial \BB_2 \cap \partial \BB_3$. Then the curvature measures of $\BB_1\cup \BB_2 \cup \BB_3$ above $p$ are given by:
$$
\Phi^H_{\BB_1\cup \BB_2\cup \BB_3}(\{p\})=0\
$$
and\
{\small
$$
\Phi^G_{\BB_1\cup \BB_2\cup \BB_3}(\{p\})
=4\ \arctan\sqrt{
\tan \left(\frac{\sigma }{2} \right)
\tan \left(\frac{\sigma - \alpha_{1,2}}{2} \right)
\tan \left(\frac{\sigma - \alpha_{2,3}}{2} \right)
\tan \left(\frac{\sigma - \alpha_{1,3}}{2} \right)
},
$$
}
where
$$
\begin{array}{l}
\sigma=\frac{\alpha_{1,2}+\alpha_{2,3}+\alpha_{3,1}}{2}\\
\alpha_{i,j}=\angle\left(\overrightarrow{pA_i},\overrightarrow{pA_j}\right)=2\ arcsin\left( \frac{A_iA_j}{2r}\right),
\end{array}
$$
\end{proposition}
\begin{proof}
Since the 2-form $\omega^H$ is mixed and since the support of the normal cycle of  $\BB_1\cup \BB_2\cup \BB_3$ ``above" $p$ lies in $\{p\}\times \R^3$, one has $\Phi^H_{\BB_1\cup \BB_2}(\{p\})=0$.
Since the normal cycle  is additive and since $\{p\}$ is 0-dimensional, by using Proposition \ref{proposition-mesures-2-boules}, one has:
$$
\begin{array}{rrl}
\Phi^G_{\BB_1\cup \BB_2\cup \BB_3}(\{p\})
&=&\Phi^G_{\BB_1}(\{p\})+\Phi^G_{\BB_2}(\{p\})+\Phi^G_{\BB_3}(\{p\})\\
&&-\Phi^G_{\BB_1\cap \BB_2}(\{p\})-\Phi^G_{\BB_1\cap \BB_3}(\{p\})-\Phi^G_{\BB_2\cap \BB_3}(\{p\})\\
&&+\Phi^G_{\BB_1\cap \BB_2\cap \BB_3}(\{p\}).\\
&=&\Phi^G_{\BB_1\cap \BB_2\cap \BB_3}(\{p\}).
\end{array}
$$
$\Phi^G_{\BB_1\cap \BB_2\cap \BB_3}(\{p\})$ is just the area of the set:
$$
S(p)=\left\{(p,\xi),\ \|\xi\|=1\ and\ \forall q \in \BB_1\cap \BB_2\cap \BB_3\ \overrightarrow{pq}.\xi \leq 0\right\}.
$$
The set $S(p)$ is a spherical triangle whose area is given by the following formulae (see \cite{berger} page 289): 
{\small
$$
4\ \arctan\sqrt{
\tan \left(\frac{\sigma }{2} \right)
\tan \left(\frac{\sigma - \alpha_{1,2}}{2} \right)
\tan \left(\frac{\sigma - \alpha_{2,3}}{2} \right)
\tan \left(\frac{\sigma - \alpha_{1,3}}{2} \right)
}.
$$
}\end{proof}

\begin{figure}[h!]
\centerline{
\includegraphics[height=6cm]{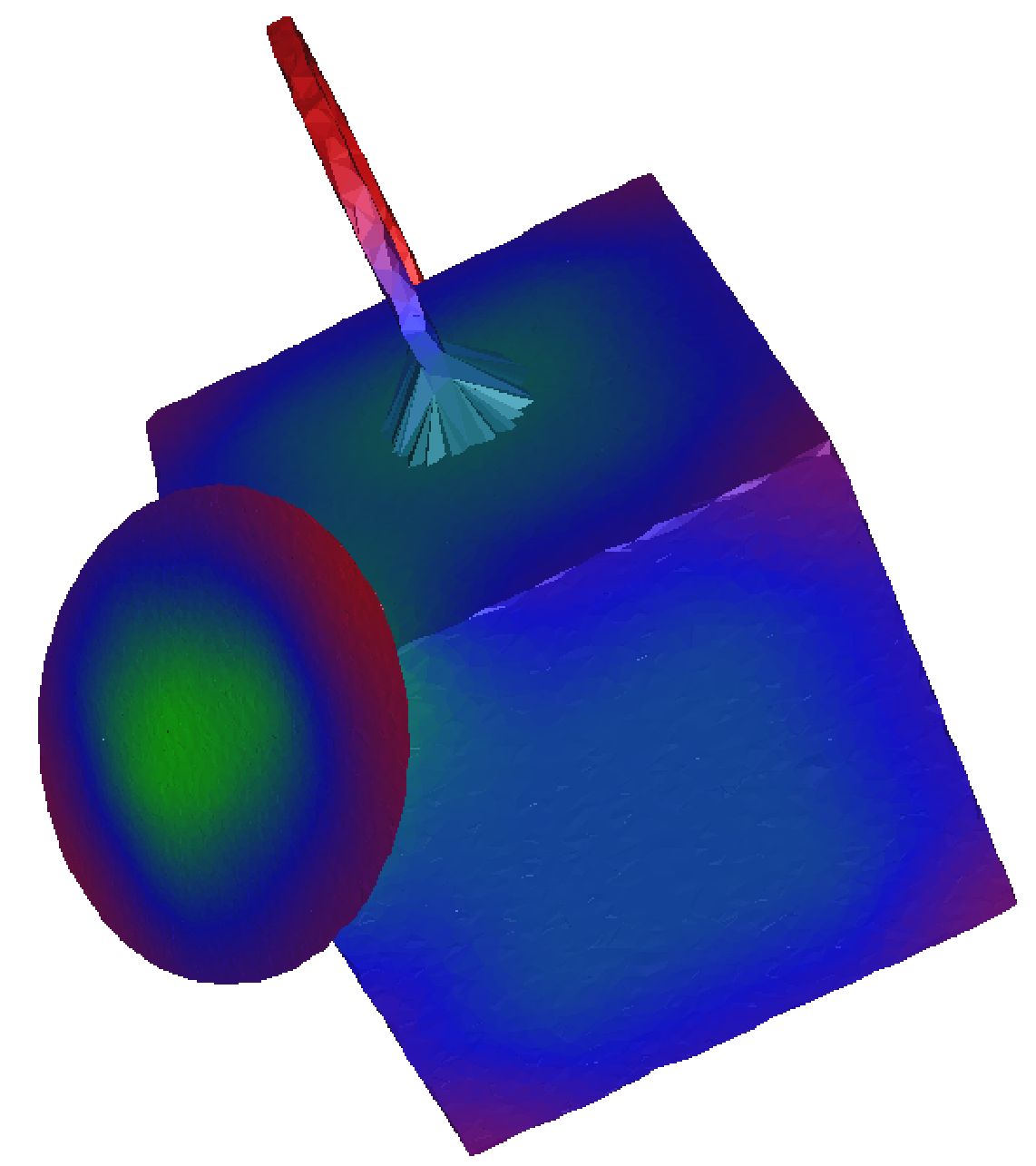}
\hspace{.8cm}
\includegraphics[height=6cm]{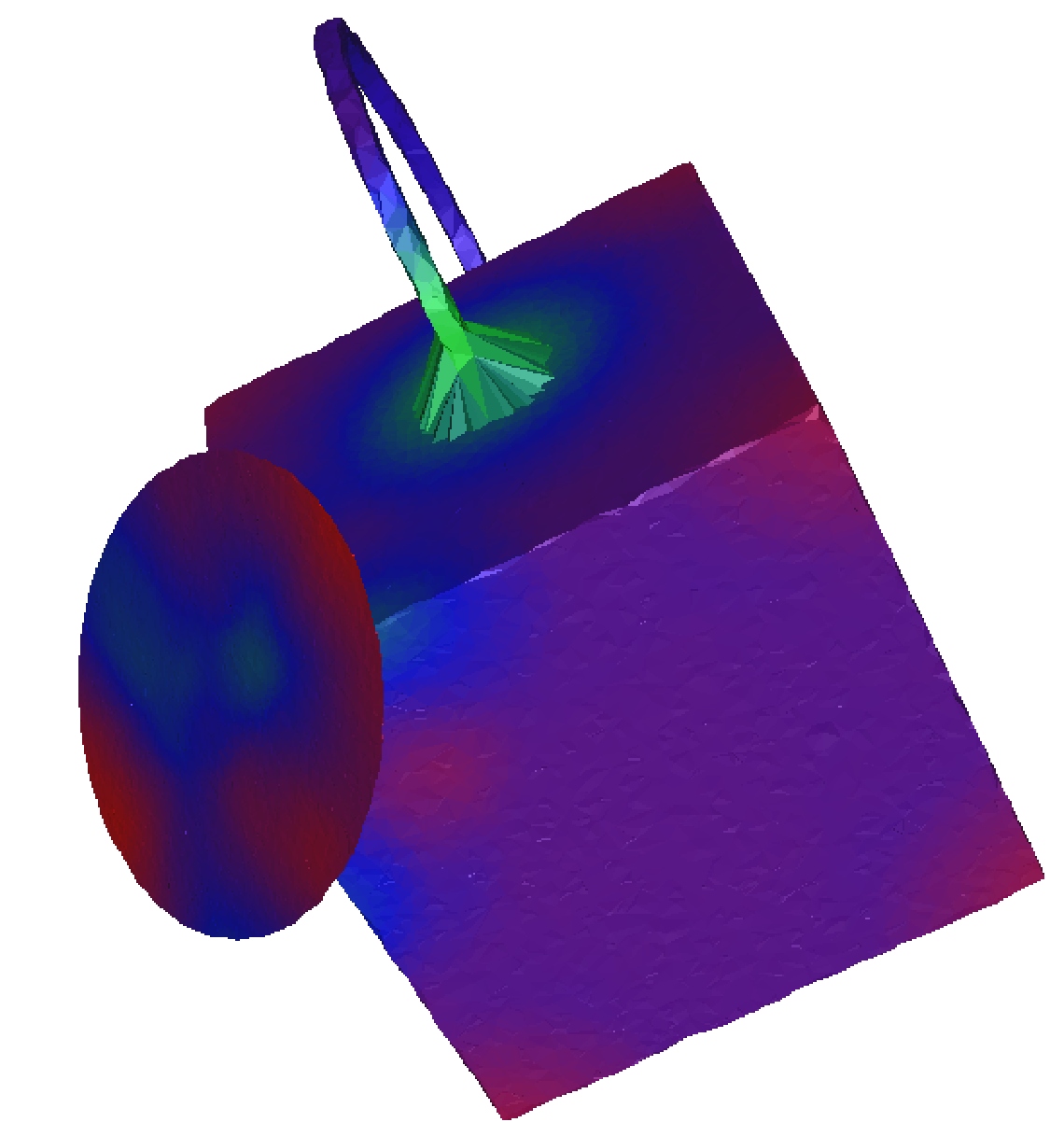}}
\centerline{
\includegraphics[height=6cm]{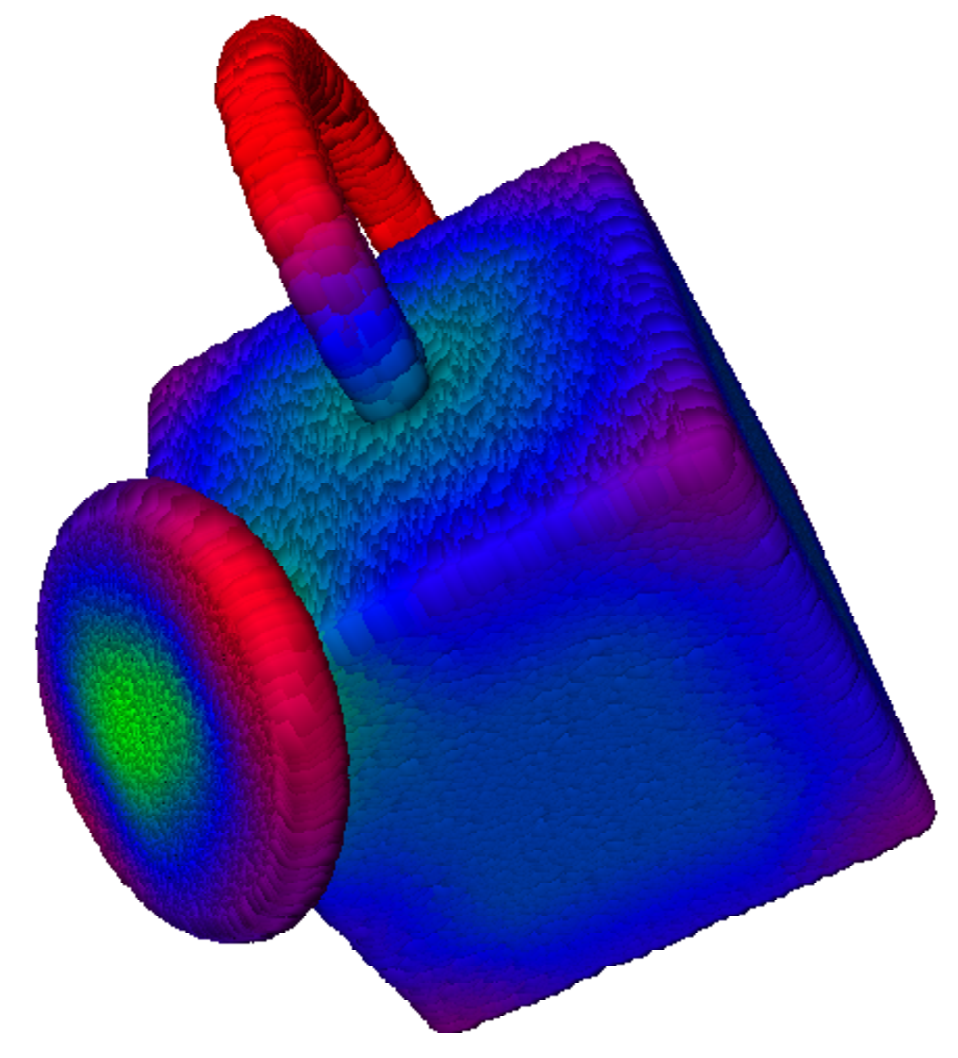}
\hspace{.8cm}
\includegraphics[height=6cm]{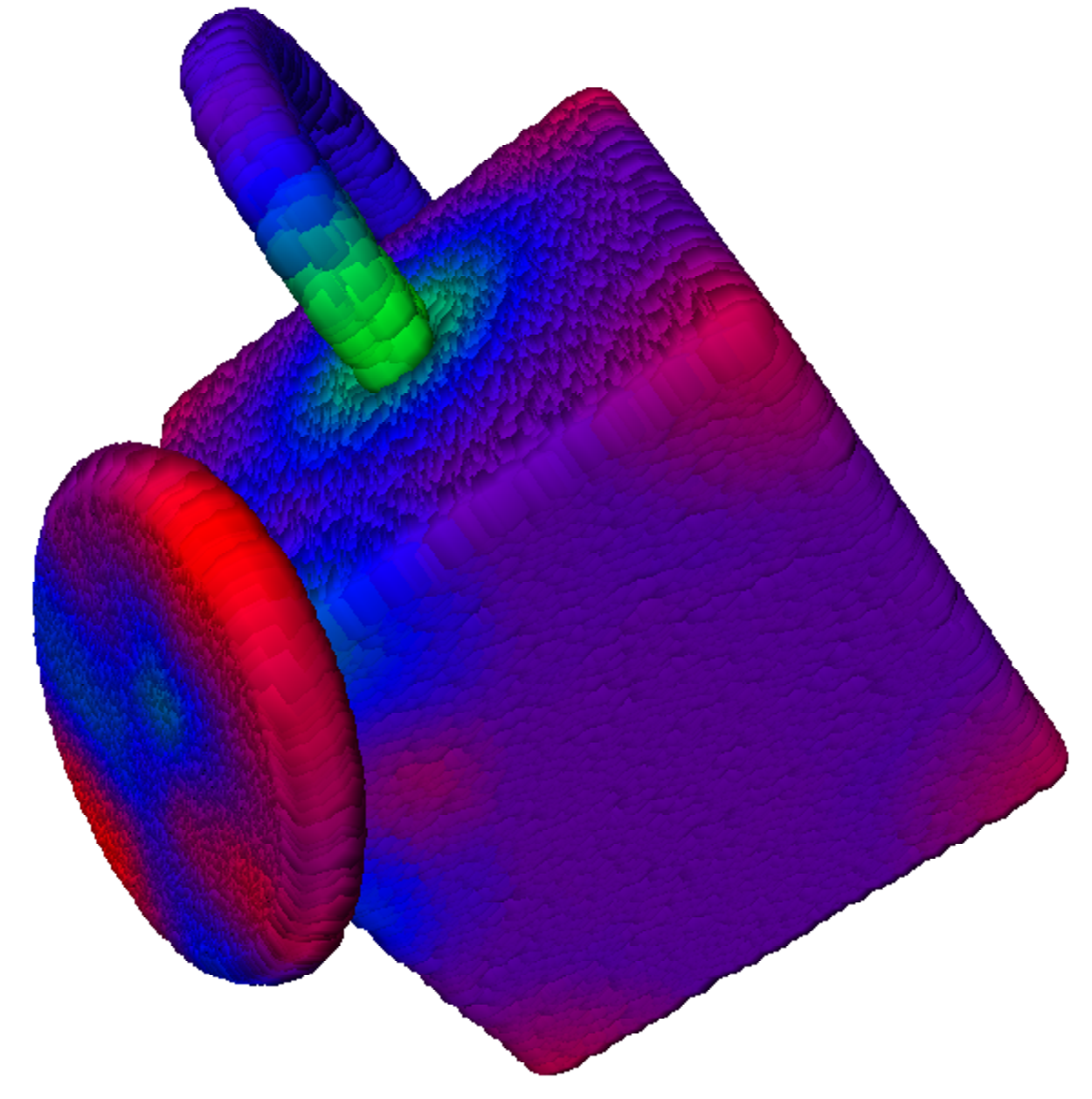}}
\caption{\small The mean (left) and gaussian (right) curvatures of a point cloud $P$ sampling a non smooth compact set union of a solid cube with a disc and an arc circle. The colors vary from green (minimum value of the curvature) to red (maximum value of the curvature) passing through blue. The curvature measures are computed using our approach for a fixed offset value of $\alpha = 0.1$ (the diameter of the point cloud being equal to $2$). A curvature value is assigned to each data point $p$ by integration of a Lipschitz function with support contained in a ball of center $p$ and radius $0.3$. These values are then used to color the faces of the boundary of the $\alpha$-shape of $P$ (top) or the offset of the point cloud (bottom).}
\label{figure-images-gauss-mean-non-manifold}
\end{figure}

\begin{figure}[h!]
\centerline{
\includegraphics[height=6cm]{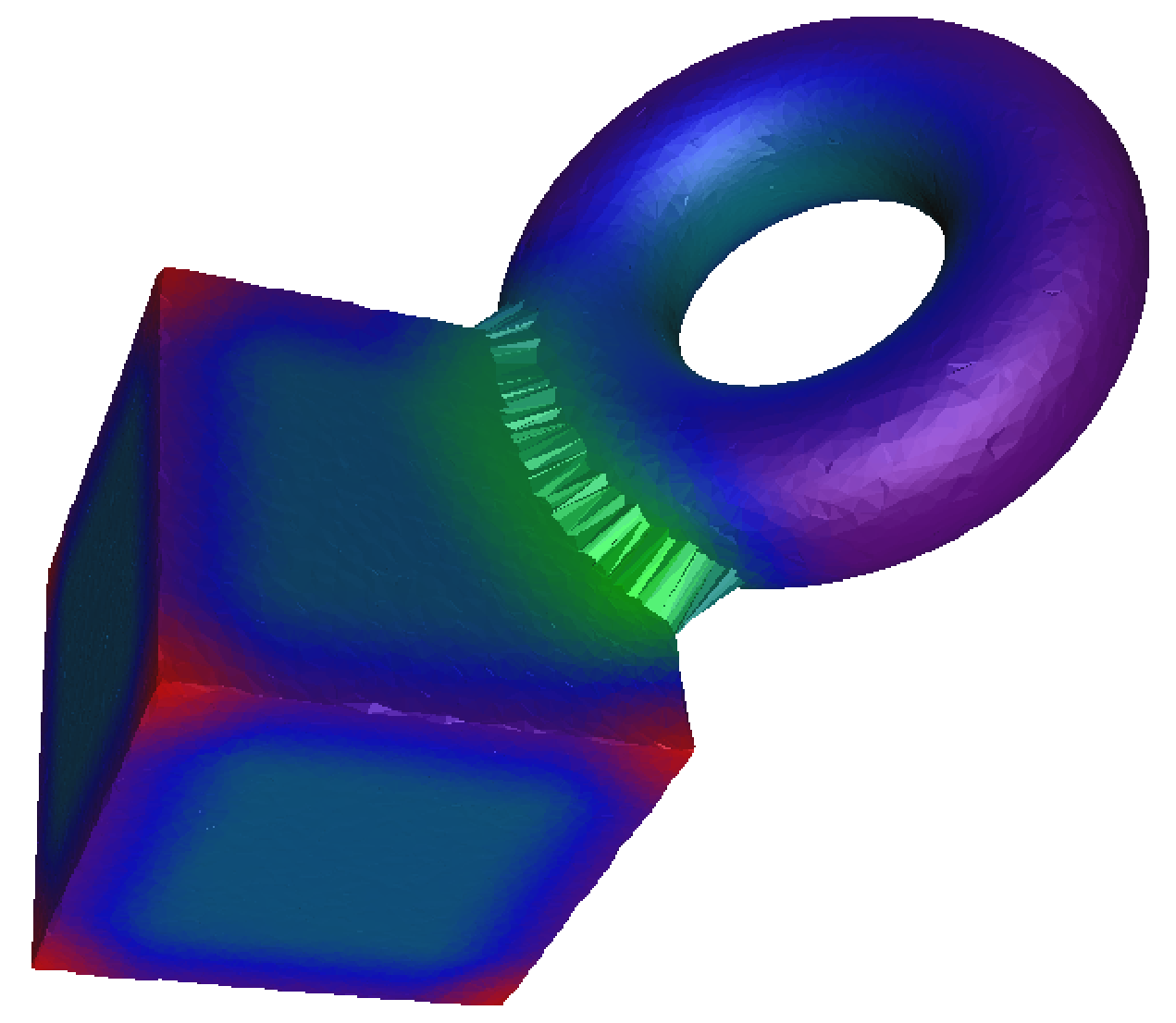}
\hspace{.5cm}
\includegraphics[height=6cm]{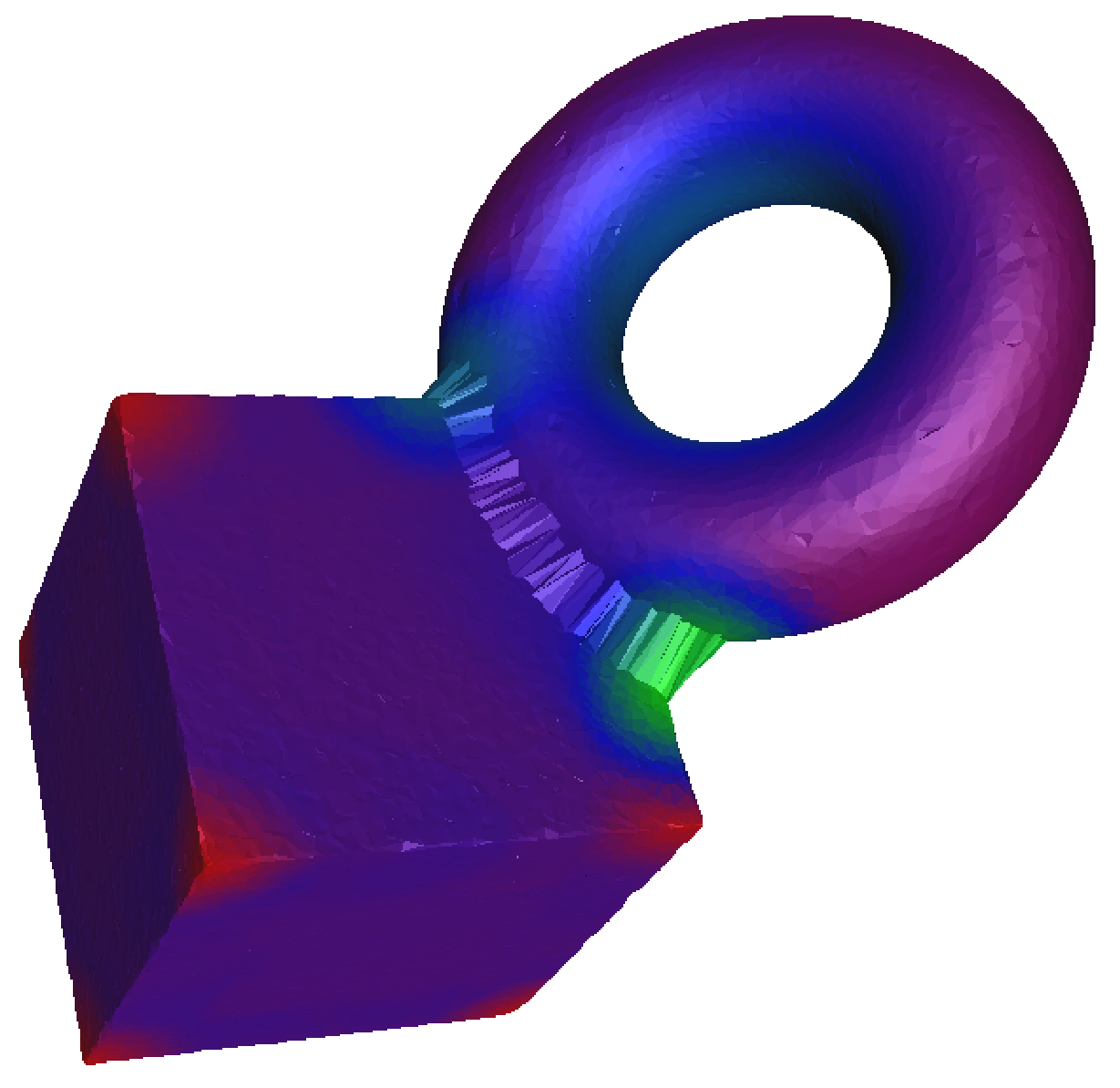}}
\caption{\small The mean (left) and gaussian (right) curvatures of a point cloud $P$ sampling the union of a solid cube and a solid torus. The colors vary from green (minimum value of the curvature) to red (maximum value of the curvature) passing through blue. The curvature measures are computed using our approach for a fixed offset value of $\alpha = 0.1$ (the diameter of the point cloud being equal to $2$). A curvature value is assigned to each data point $p$ by integration of a Lipschitz function with support contained in a ball of center $p$ and radius $0.3$. These values are then used to color the faces of the boundary of the $\alpha$-shape of $P$.}
\label{figure-images-gauss-mean}
\end{figure}

In Figure \ref{figure-images-gauss-mean-non-manifold}, we sampled a non-manifold compact set. It is interesting to note that both the mean and the gaussian curvature are low (and negative) in the middle of the disc. This is due to the fact that the curvature is ``caught" by a function $f$ whose support traverses the disc and intersects the cube. 
Figure \ref{figure-images-gauss-mean} also illustrates the algorithm for a sampling of the union of a torus and a cube. 

\noindent {\bf The algorithm}\\
In practice, we use a halfedge data structure (provided by S. Loriot from INRIA Sophia-Antipolis) for the representation of the boundary of the union of balls $K_r$. Let now $f$  be a Lipschitz function on $\R^d$. To compute the isotropic curvature measures $\Phi^G_{K_r}(f)$ and $\Phi^H_{K_r}(f)$ of the union of balls $K_r$, we triangulate  $\partial K_r$. More precisely, we triangulate every face of $\partial K_r$ into ``small" spherical patches of diameter less than $\eta$. We then approximate $f$ by a function $\tilde f$ that is constant on each ``small" patch and that coincides with $f$ in at least one point of each patch. We then calculate the isotropic curvature measures with $\tilde f$ ``above the faces" by using Proposition \ref{proposition_1ball}. Similarly, by using Proposition \ref{proposition-mesures-2-boules}, we calculate the isotropic curvature measures above the ``spherical edges"  with a function $\tilde f$ piecewise constant on ``small" edges of lengths less than $\eta$. Finally, we calculate exactly the isotropic curvature measures ``above" the vertices by using Proposition \ref{proposition-mesures-3-boules}. 

\noindent {\bf Numerical integration}\\
Let $B'$ be a borel subset of $\R^3$ that contains the support of the functions $f$ and let $B=B'\cap \partial K_r$. The numerical error done by approximating $f$ by a piecewise constant function $\tilde f$ is then given by:
$$
|\Phi^G_{K_r}(f) - \Phi^G_{K_r}(\tilde{f})|
 \leq Lip(f)\  M(N(K_r)\llcorner (B\times \R^d))\ \eta
$$
and
$$
|\Phi^H_{K_r}(f) - \Phi^H_{K_r}(\tilde{f})|
 \leq Lip(f)\  M(N(K_r)\llcorner (B\times \R^d))\ \eta
$$
where $\tilde f$ is constant on each patch and coincides with $f$ at at least one point in the patch.

\noindent {\bf Proof:} We show the result for the Gaussian curvature measure. The proof for the mean curvature measure is similar. One has:
$$
\begin{array}{rl}
|\Phi^G_{K_r}(f) - \Phi^G_{K_r}(\tilde{f})|
&\ds = \int_{spt(N(K_r))\cap (B\times \R^d)} (\bar{f}-\bar{\tilde{f}})\ \omega^G\\
&\ds \leq \sup_{(p,n)\in (B\times \R^d)}\|\bar{f}(p,n)-\bar{\tilde{f}}(p,n)\|\ M(N(K_r)\llcorner (B\times \R^d)).
\end{array}
$$
Let $F$ be a face of the spherical polygon $\partial K_r \cap B'$, that is decomposed into spherical patches $\Delta$. Let  $\tilde f$ be the piecewise constant function equal to $f(p_0)$ on every spherical patch $\Delta$ of $F$ (with $p_0\in \Delta$). We then have:
$$
\sup_{(p,n)\in (F\times \R^d)}\|\bar{f}(p,n)-\bar{\tilde{f}}(p,n)\|
\leq Lip(f)\ \sup_{\Delta\subset F}\eta_{\Delta},
$$ where $\eta_{\Delta}$ is the length of the longest edge of $\Delta$. The result is similar for a spherical edge $C$ of $\partial K_r \cap B'$: if one decomposes $C$ into small arc of circles $e$ one has:
$$
\sup_{(p,n)\in (C\times \R^d)}\|\bar{f}(p,n)-\bar{\tilde{f}}(p,n)\|
\leq Lip(f)\ \sup_{e \subset C} l(e),
$$
where $l(e)$ is the length of the arc of circle $e$.

\subsection{Anisotropic curvature measures of $3D$ point clouds}
In the previous section, we described an algorithm that allows to calculate the curvature measures  $\Phi^G_{K_r}(f)$  and  $\Phi^H_{K_r}(f)$ when $K$ is a finite set of points. Similarly, one can also calculate the anisotropic curvature measure $\overline{H}_{K_r}(f)$ and $\overline{\widetilde{H}}_{K_r}(f)$ of $K_r$ by using the curvature measures formulae for the union of one, two or three balls (see Propositions \ref{proposition-anistrope-1ball}, \ref{proposition-anisotrope-2-boules} and \ref{proposition-anistrope-3ball}). More details on how to use these formulae can be found in \cite{cohen-steiner,cohen-steiner-morvan-socg03}.

\begin{proposition}\label{proposition-anistrope-1ball}
Let $\BB$ be a ball of $\R^3$ and let $B$ be a Borel set of $\R^3$. Then the anisotropic curvature measures of $\BB$ above $B$ are given by the $3\times 3$-matrix:
$$
\overline{\widetilde{H}}_{\BB}(B)=\overline{H}_{\BB}(B) = \int_{B\cap \partial \BB} \frac{1}{r}\ (Id - n_p {}^tn_p)\ dp, 
$$
where $Id$ is the identity matrix of $\R^3$, $n_p$ is the unitary normal of $\BB$ at $p$ and 
${}^tn_p$ its transpose.
\end{proposition}
\begin{proof}
Let $X,Y$ be two vectors of $\R^3$ and $P_p$ denotes the projection onto the plane tangent to $\BB$ at the point $p$. Using that $P_p(X)=X-<X,n_p>n_p$, we have:
$$
\overline{H}_{\BB}(B)(X,Y)=\int_{B\cap \partial \BB} H_p(P_p(X),P_p(Y))dp
=\int_{B\cap \partial \BB} \frac{1}{r}\ (Id - n_p {}^tn_p)\ dp.
$$
\end{proof}

\begin{proposition}\label{proposition-anisotrope-2-boules}
Let $\BB_1$ and $\BB_2$ be two intersecting balls of $\R^3$, of same radius $r>0$ and   of centers $A_1$ and  $A_2$, and $C$ be the circle $\partial \BB_1 \cap \partial \BB_2$.  Let $B$ be a ball of $\R^3$. Then the anisotropic curvature measures of $\BB_1\cup \BB_2$ above $B\cap C$ are given by the $3\times 3$ matrix $\overline{H}_{\BB_1\cup \BB_2}(B\cap C)$:
$$
-r \sin \alpha 
\left(\begin{array}{ccc}
\frac{1}{2} \left(\alpha - \frac{\sin 2 \alpha}{2}\right) \left(\beta  + \frac{\sin 2 \beta}{2}\right)
&\left(\alpha - \frac{\sin 2 \alpha}{2}\right) \frac{1 - \cos 2 \beta}{4}
&0\\
\left(\alpha - \frac{\sin 2 \alpha}{2}\right) \frac{1 - \cos 2 \beta}{4}
&\frac{1}{2} \left(\alpha - \frac{\sin 2 \alpha}{2}\right) \left(\beta  - \frac{\sin 2 \beta}{2}\right)
&0\\
0&0
&\beta \left(\alpha + \frac{\sin 2 \alpha}{2}\right)
\end{array}\right),
$$
and the $3\times 3$ matrix $\overline{\widetilde{H}}_{\BB_1\cup \BB_2}(B\cap C)$:
$$
r \alpha \sin \alpha 
\left(\begin{array}{ccc}
 \left(\beta  - \frac{\sin 2 \beta}{2}\right)
& \frac{ \cos 2 \beta - 1}{2}
&0\\
\frac{ \cos 2 \beta - 1}{2}
& \left(\beta  - \frac{\sin 2 \beta}{2}\right)
&0\\
0&0
&0
\end{array}\right),
$$
in the orthonormal frame $(0,i,j,k)$ where $(0,i,j)$ generates the plane of the circle $C$ and $i$ points toward one on the two extremities of the arc of circle $B\cap C$,
where $\alpha$ and $\beta$ are defined as in the proof of Proposition \ref{proposition-mesures-2-boules}.
\end{proposition}
\begin{proof}
We take the same notations that the ones of Proposition \ref{proposition-mesures-2-boules}. We have:
$$
\overline{H}_{\BB_1\cup \BB_2}(B\cap C)(X,Y)
=\int_0^{\beta}\int_{-\alpha}^{\alpha}f^*\omega^{X,Y}.
$$
As in Proposition \ref{proposition-mesures-2-boules}, we take $(m,\xi)=f(u,v) \in S_C(\BB_1 \cap \BB_2)$. We have:
$$
\begin{array}{rl}
f^*\omega^{X,Y}((1,0),(0,1))
&=\omega^{X,Y}_{(m,\xi)}\left(\frac{\partial f}{\partial u}(u,v),\frac{\partial f}{\partial v}(u,v) \right)\\
&=(0,Y)\ \wedge (\xi\times X,0)((r \sin \alpha\ e_1,\cos v\ e_1),(0,e_2))\\
&=\left|\begin{array}{cc}
(0,Y).(r \sin \alpha\ e_1,\cos v\ e_1) &(\xi\times X,0).(r \sin \alpha\ e_1,\cos v\ e_1)\\
(0,Y).(0,e_2) &(\xi\times X,0).(0,e_2)\\
\end{array}\right|\\
&=\left|\begin{array}{cc}
\cos v\ Y.e_1& \det(\xi, X,r\sin \alpha e_1)\\
Y.e_2&0\\
\end{array}\right|\\
&=-r\sin \alpha\ X.e_2\ Y.e_2.\\
\end{array}
$$
The result is thus obtained by integrating  the matrix of $e_2{}^te_2$ given by:
$$
\left(\begin{array}{ccc}
\sin^2v \cos^2u &\sin^2v \sin u \cos u & -\sin v \cos v \cos u\\
\sin^2v \sin u \cos u &\sin^2v\sin^2u & -\sin v \sin u \cos v\\
-\cos v \sin v \cos u & -\cos v \sin v \sin u & \cos^2 v\\
\end{array}\right).
$$
Similarly, we have:
$$
f^*\widetilde{\omega}^{X,Y}((1,0),(0,1))
=r\sin \alpha\ X.e_1\ Y.e_1.
$$
The result for $\overline{\widetilde{H}}_{\BB_1\cup \BB_2}(B\cap C)$ is thus obtained by integrating  the matrix of $e_1{}^te_1$ given by:
$$
\left(\begin{array}{ccc}
\sin^2u  & -\sin u \cos u & 0\\
- \sin u \cos u &\cos^2u & 0\\
0&0&0\\
\end{array}\right).
$$
\end{proof}

\begin{proposition}\label{proposition-anistrope-3ball}
Let $\BB_1$, $\BB_2$ and $\BB_3$ be three intersecting balls of $\R^3$, of radius $r$,  of centers $A_1$, $A_2$ and $A_3$, and $p\in \partial \BB_1\cap  \partial \BB_2 \cap \partial \BB_3$. Then the anisotropic curvature measures of $\BB_1\cup \BB_2 \cup \BB_3$ above $p$ are equal to $0$:
$$
\overline{H}_{\BB_1\cup \BB_2\cup \BB_3}(\{p\})=0\
and\
\overline{\widetilde{H}}_{\BB_1\cup \BB_2\cup \BB_3}(\{p\})=0\
$$
\end{proposition}
\begin{proof}
The two forms $\omega^{X,Y}$ and $\widetilde{\omega}^{X,Y}$ being mixed, they vanish on the normal component.
\end{proof}

\section{About the normal cycle of a compact of $\R^d$}\label{section-one-compact}
In this section, we give preliminary results for compact sets in $\R^d$: in Lemma \ref{lemma-mass-surface-c1}, we bound the mass of the normal cycle of a set whose boundary is of class $\C^1$; in Lemma \ref{lemma-Zahle}, we recall a result of Z\"{a}hle et al. that allows to define, under several assumptions, the normal cycle of a set whose complement has positive reach; in Lemma \ref{lemma-erosions}, we give relationships between the normal cycle of  a set whose complement has positive reach and the normal cycle of its offset.

We first give the following lemma (that is a direct consequence of the Isotopy Lemma, Proposition 1.8 in \cite{grove}):
\begin{lemma}\label{lemma-critical}
Let $K$ be a compact of $\R^d$. If $r$ is not a critical point of the distance function ({\it i.e.} $\chi_K(r)\neq 0$), we then have:
$$
\overline{\{x,\ d_K(x)<r\}} = \{x,\ d_K(x)\leq r\}
\quad \mbox{and} \quad
\overline{\{x,\ d_K(x)>r\}} = \{x,\ d_K(x)\geq r\}.
$$
\end{lemma}

\begin{lemma}\label{lemma-mass-surface-c1}
Let $X$ be a compact subset of $\R^d$ whose boundary is a hypersurface with positive reach $R$. We then have:
$$
M(N(X)) \leq \left(1+ \frac{1}{R^2}\right)^{\frac{d-1}{2}}\ \H^{d-1}(\partial X).
$$
More precisely, if $B'$ is a Borel of $\R^d$ and $B=B'\cap \partial X$, we have:
$$
M\left[N(X)\llcorner (B\times \R^d)\right] \leq \left(1+ \frac{1}{R^2}\right)^{\frac{d-1}{2}}\ \H^{d-1}(B).
$$
\end{lemma}

\begin{proof}
Let us consider the following map:
$$
\begin{array}{rclc}
f: & B &\to &spt(N(X))\cap (B\times \R^d)\\
&m &\mapsto &\left(m,n(m)\right)
\end{array}.
$$

Since the reach, and therefore the curvature of the  surface $\partial X$ is bounded by $R$, it follows that
the map $m \mapsto n(m)$ is $\frac{1}{R}$-Lipschitz. That implies that $f$ is also $\sqrt{1+ \frac{1}{R^2}}$-Lipschitz. Since the map $f$ is one-to-one, the general coarea formulae (see Theorem 4.2 of \cite{morgan-ritore} or Proposition 3.13 of \cite{morgan}) states that: 
$$
\begin{array}{rl}
\displaystyle M(N(X)\llcorner (B\times \R^d)) 
&\displaystyle =\int_{spt(N(X))\cap (B\times \R^d)}d\H^{d-1}\\
& \displaystyle =\int_{B}J_{d-1}(f)\ d\H^{d-1}\\
& \displaystyle\leq \int_{B}\ \left(1+ \frac{1}{R^2}\right)^{\frac{d-1}{2}}\ d\H^{d-1}\\
 & \displaystyle\leq \left(1+ \frac{1}{R^2}\right)^{\frac{d-1}{2}}\ \H^{d-1}(B). 
\end{array}
$$
where $J_{d-1}(f)$ is the (d-1)-jacobian of the Lipschitz map $f$ (see \cite{morgan}, page 24-25).
\end{proof}

Under certain assumptions, Rataj and Z\"{a}hle \cite{rataj-Zahle-03} showed that the closure of the complement of a compact set with positive reach admits a normal cycle. 
More precisely, we denote by $i$ the inversion map 
$$
\begin{array}{rclc}
i: &\R^d \times \S^{d-1} &\to &\R^d \times \S^{d-1}\\
&(x,n) &\mapsto &(x,-n),\\
\end{array}
$$
and by $i_{\sharp}$ the push-forward by $i$. We define the erosions $X_{-\epsilon}$ of $X$ by $X_{-\epsilon}=\overline{\left(\overline{X^c}_{\, \epsilon}\right)^c}$. We then get the following lemma by combining results of Z\"{a}hle et al. \cite{rataj-Zahle-01,rataj-Zahle-03} and \cite{doubleOffset}:
\begin{lemma}\label{lemma-Zahle}
Let $K$ be a compact set of $\R^d$ with $\mu$-reach greater than $r$ and let $X=K_r$. We can then define the normal cycle of $X$ as the flat limit of its erosions as follows:
$$
\F\left( N\left(X_{-\epsilon}\right) - N(X) \right) \longrightarrow_{\epsilon \to 0} 0.
$$
Furthermore, since $\overline{X^c}$ has positive reach, it has a normal cycle and we have:
$$
N(X)=-i_{\sharp} N\left( \overline{X^c} \right)
\quad \mbox{and} \quad
M(N(X))=M\left( N\left( \overline{X^c} \right)\right).
$$
\end{lemma}
\begin{proof}
Proposition 2 and Theorem 3 of \cite{rataj-Zahle-03} imply that the normal cycle $N(X)$ of $X$ is well defined and we have:
$$
\F\left( N\left(X_{-\epsilon}\right) - N(X) \right) \longrightarrow_{\epsilon \to 0} 0.
$$
Now, since $\overline{X^c}$ has positive reach \cite{doubleOffset}, Corollary 3.1 of \cite{rataj-Zahle-01} (or Proposition 4 of \cite{rataj-Zahle-03}) implies that:
$$
\F\left[
N\left( \overline{X^c} \right) - \left(-i_{\sharp} N\left(X_{-\epsilon}\right) \right) 
\right] \longrightarrow_{\epsilon \to 0} 0.
$$
We then have
$$
N(X)=-i_{\sharp} N\left( \overline{X^c} \right).
$$
\end{proof}

We now consider the following map
$$
\begin{array}{cccc}
F_t :&\R^d\times\R^d&\to &\R^d\times \R^d\\
&(x,v)&\mapsto&(x+tv,v)\\
\end{array},
$$
and denote by  $F_{t \sharp}$ the associated push-forward for currents.

\begin{lemma}\label{lemma-erosions}
Let $V$ be a compact set of $\R^d$ with positive reach $R>0$. We then have for $\epsilon < R$:
$$
N\left(V\right)
= F_{-\epsilon \sharp} \left[N\left( V_{\, \epsilon} \right)\right].
$$
That result can be localized. Let $B'$ be a Borel of $\R^d$ and $B=(B'\times \R^d)\cap spt(N(V))$. We then have:
$$
N\left( V \right) \llcorner B
= F_{-\epsilon \sharp} \left[N\left( V_{\, \epsilon} \right)\llcorner F_{-\epsilon}^{-1}(B)\right],
$$

In particular, if $K$ is a compact set of $\R^d$ with $\mu$-reach greater than $r$, $X=K_r$, the set $V=\overline{X^c}$ has a reach $R>\mu r$, and we have for $\epsilon < R$:
$$
\begin{array}{rl}
&N\left( X \right) 
= -i_{\sharp} N\left( \overline{X^c} \right)
= -i_{\sharp} F_{-\epsilon \sharp} \left[N\left( \overline{X^c}_{\, \epsilon}  \right)\right]\\
\mbox{and} 
&M\left( N\left( X \right) \right)
\leq \left(1+\epsilon^2\right)^{\frac{d-1}{2}}\ M\left[N\left( \overline{X^c}_{\, \epsilon}  \right)\right].\\
\end{array}
$$
Let now $B'$ be a Borel of $\R^d$ and $B=(B'\times \R^d)\cap spt(N(X))$. We then have:
$$
N\left( X \right) \llcorner B
= - i_{\sharp}F_{-\epsilon \sharp} \left[N\left( \overline{X^c}_{\, \epsilon}  \right) \llcorner F_{-\epsilon}^{-1}(i (B))\right],
$$
$$
M\left( N\left( X \right) \llcorner B\right)
\leq \left(1+\epsilon^2\right)^{\frac{d-1}{2}}\ M \left[N\left( \overline{X^c}_{\, \epsilon}  \right) \llcorner F_{-\epsilon}^{-1}(i(B))\right].
$$
\end{lemma}

\begin{proof}
We notice that  the restriction of $F_{-\epsilon}$ to the support of $N\left(V_{\, \epsilon}\right)$ is a one-to-one map:
$$
\begin{array}{cccc}
F_{-\epsilon} :&spt\left( N\left(V_{\, \epsilon}\right)\right)&\to &spt\left( N\left(V\right)\right)\\
&(x,v)&\mapsto&(x-\epsilon v,v)\\
\end{array}.
$$
The general coarea formula implies that:
$$
M\left(N\left( V \right) \right)
=\int_{spt\left( N\left(V\right)\right)} d\H^{d-1}
=\int_{spt\left( N\left(V_{\, \epsilon}\right)\right)} J_{d-1}(F_{-\epsilon})\ d\H^{d-1}.
$$
The map $F_{-\epsilon}$ is $\sqrt{1+\epsilon^2}$-Lipschitz, thus we have $J_{d-1}(F_{-\epsilon}) \leq \left(1+\epsilon^2\right)^{\frac{d-1}{2}}$ and:
$$
M\left( N\left( V \right)\right)
\leq \left(1+\epsilon^2\right)^{\frac{d-1}{2}}\ 
M\left( N\left(V_{\epsilon}\right)\right).
$$
More precisely, for any $d-1$-differential form $\varphi$ on $\R^d\times \R^d$, one has:$$
\begin{array}{rl}
F_{-\epsilon \sharp} N\left( V_{\, \epsilon}\ \right) (\varphi)
&= N\left( V_{\, \epsilon}\ \right) \left(F_{-\epsilon}^* \varphi \right)\\
&\displaystyle = \int_{spt\left( N\left(V_{\, \epsilon}\right)\right)} (F_{-\epsilon}^*\varphi)_x (e_1^x\wedge...\wedge e_{d-1}^x)\ d\H^{d-1}(x)\\
&\displaystyle = \int_{spt\left( N\left(V_{\, \epsilon}\right)\right)}  \varphi_{F_{-\epsilon}(x)}((DF_{-\epsilon}(x)e_1^x)\wedge...\wedge (DF_{-\epsilon}(x)e_{d-1}^x))\ d\H^{d-1}(x)\\
&\displaystyle = \int_{spt\left( N\left(V_{\, \epsilon}\right)\right)}  J_{d-1}(F_{-\epsilon})(x)\ \varphi_{F_{-\epsilon}(x)}\left(\frac{(DF_{-\epsilon}(x)e_1^x)\wedge...\wedge (DF_{-\epsilon}(x)e_{d-1}^x)}{|(DF_{-\epsilon}(x)e_1^x)\wedge...\wedge (DF_{-\epsilon}(x)e_{d-1}^x)|}\right)\ d\H^{d-1}(x)\\
&\displaystyle = \int_{spt\left( N\left(V_{\, \epsilon}\right)\right)}  J_{d-1}(F_{-\epsilon})(x)\ \varphi_{F_{-\epsilon}(x)}\left(  
e_1^{F_{-\epsilon}(x)}\wedge...\wedge e_{d-1}^{F_{-\epsilon}(x)}
\right)\ d\H^{d-1}(x)\\
&\displaystyle = \int_{spt\left( N\left( V \right)\right)}  \varphi_{y}\left(  
e_1^{y}\wedge...\wedge e_{d-1}^{y}
\right)\ d\H^{d-1}(y)\quad \mbox{\ (by the general coarea formula)}\\
&\displaystyle =N\left( V \right) \left( \varphi \right),\\
\end{array}
$$
where $e_1^x\wedge..\wedge e_{d-1}^x$ is a unit (d-1)-vector associated with the oriented tangent space $T_x\left(spt\left( N\left(V_{\, \epsilon}\right)\right)\right)$ and $e_1^y\wedge..\wedge e_{d-1}^y$ is a unit (d-1)-vector associated with the oriented tangent space $T_y\left(spt\left( N\left( V \right)\right)\right)$ and $J_{d-1}(F_{-\epsilon})(x)=|(DF_{-\epsilon}(x)e_1^x)\wedge...\wedge (DF_{-\epsilon}(x)e_{d-1}^x)|$ $\H^{d-1}$-almost everywhere.
That implies that:
$$
N\left( V \right) 
= F_{-\epsilon \sharp} N\left( V_{\, \epsilon} \right).
$$
The local version is done the same way by noticing that the restriction of $F_{-\epsilon}$ to the support of $N\left(V_{\, \epsilon}\right)\llcorner F_{-\epsilon}^{-1}(B)$ is a one-to-one map:
$$
\begin{array}{cccc}
F_{-\epsilon} :&spt\left( N\left(V_{\, \epsilon}\right)\llcorner F_{-\epsilon}^{-1}(B)\right)&\to &spt\left( N\left(V\right)\llcorner B \right)\\
&(x,v)&\mapsto&(x-\epsilon v,v)\\
\end{array}.
$$
\end{proof}

\section{Normal cycles of the double offsets of two close compact sets}\label{section-double-offsets}
The aim of this section is to prove the following proposition. It states that if two compact sets $K$ and $K'$ are close in the Hausdorff sense, then the normal cycles of their double offsets are close for the flat norm. This result is local since we evaluate the normal cycles locally.

More precisely, let $B'$ be a Borel subset of $\R^d$ and $B=B'\cap \partial K'_{r,t}$. We consider in the following the normal cycles of $K_{r,t}$ and $K'_{r,t}$ restricted to $B$ and $\pi(B)$, where $\pi$ is the projection onto $K_{r,t}$. We consider the two currents $D'=N(K'_{r,t})\llcorner (B\times \R^d)$ and $D=N(K_{r,t})\llcorner (\pi(B)\times \R^d)$ given by: 
$$
\begin{array}{rl}
&\displaystyle D'(\varphi)=\int_{spt(N(K'_{r,t}))\cap (B\times \R^d)} \varphi\\ 
\mbox{and} 
& \displaystyle D(\varphi)=\int_{spt(N(K_{r,t}))\cap (\pi(B)\times \R^d)} \varphi,
\end{array}
$$
for any (d-1)-differential form $\varphi$ of $\R^d\times\R^d$.

\begin{proposition}\label{proposition-double-offsets}
Let $K$ and $K'$ be two compact sets of $\R^d$ whose $\mu$-reaches are greater than $r$. We suppose that the Hausdorff distance $\epsilon=d_H(K,K')$  between $K$ and $K'$ is less than $\frac{r\mu\ (2-\sqrt{2})}{2} \min(\mu,\frac{1}{2})$. For $t\in \left[\frac{\epsilon}{\min(\mu,\frac{1}{2})\ (2-\sqrt{2})}, \frac{r\mu}{2}\right]$, we can write:
$$
D' - D = \partial R_1 + R_2
$$
where $R_1$ and $R_2$ are currents satisfying
\begin{eqnarray*}
M(R_1) &\leq& M(D')  k_1k_2\\
M(R_2) &\leq& M(\partial D') k_1k_2,
\end{eqnarray*}
where
\begin{eqnarray*}
k_1 &=& 1 + \left(\frac{1+t^2}{\left(t-\frac{\epsilon}{\mu}\right)^2}\right)^{\frac{d-1}{2}}\\
k_2 &=& \sqrt{\left(\frac{\epsilon}{\mu}\right)^2 + 900\frac{\epsilon}{\mu t}}.
\end{eqnarray*}
In particular the flat norm of $D-D'$ is bounded by $(M(D')+M(\partial D'))k_1k_2$.
\end{proposition}
The proof of that proposition relies on the two following lemmas:
\begin{lemma}\label{lemma-normals}
Let $X$ and $X'$ be two compact sets  of $\R^d$ with reaches greater than $R>0$.  Let $t \leq \frac{R}{2}$ and $\epsilon=d_H(X,X')$. If $\epsilon \leq \frac{t}{2}$, then for any $x$ at a distance $t$ from $X$, we have:
$$
2 \sin \frac{\angle\left(\nabla_X(x),\nabla_{X'}(x)\right)}{2} \leq 30  \sqrt{\frac{\epsilon}{t}}.
$$
\end{lemma}
\begin{proof}
For $\rho$ such that $0< \rho < t$, we denote by $G_X( x,\rho)$,
as in  section 5 of \cite{NormalConeApproximation}, the convex hull for every $y\in B(x,\rho)$ of all the ``classical" gradients $\nabla_{X} (y)$ 
of the distance function $d_X$. By using Theorem 5.6 in \cite{NormalConeApproximation}, we have that:
\begin{equation}\label{EaquationFromNormalConeApproximation}
\nabla_{X'} (x) \in G_X( x,\rho)_{\frac{\rho}{2 d(x,X')} + \frac{2 \epsilon}{\rho}}.
\end{equation}
On another hand we know (\cite{federer-59} page 435) that the projection map $\pi_X$ on $X$ is $\frac{R}{R-(t+\rho)}$-Lipschitz for points at distance less than $(t+\rho)$ from $X$.
Then, for $y\in B( x,\rho)$ one has $d(x,y) \leq \rho$ and $d(\pi_X(x) ,\pi_X(y) ) \leq  \frac{R \rho}{R-(t+\rho)} $ and:
$$
\|  \overrightarrow {y \pi_X(y)} -  \overrightarrow {x \pi_X(x)} \| \leq \rho + \frac{R \rho}{R-(t+\rho)} .
$$
Using the fact that $\nabla_{X} (z) = \frac{-1}{\| \overrightarrow {z \pi_X(z)} \|} \overrightarrow {z \pi_X(z)} $ one get, for $y\in B( x,\rho)$:
$$
\|  \nabla_{X} (y) - \nabla_{X} (x) \| \leq \frac{1}{t- \rho} \left( \rho + \frac{R \rho}{R-(t+\rho)} \right).
$$
This and Equation(\ref{EaquationFromNormalConeApproximation}) gives:
$$
2 \sin \frac{\angle \nabla_X(x),\nabla_{X'}(x)}{2} =
\| \nabla_X(x) - \nabla_{X'}(x)\|
\leq \frac{\rho}{t-\rho} \left(1+\frac{R}{R-(t+\rho)}\right) + \frac{\rho}{2(t-\epsilon)}+ \frac{2\epsilon}{\rho}.
$$
Taking $\rho=\sqrt{\epsilon t}$ and $\epsilon \leq \frac{t}{2}$ one gets:
$$
2 \sin \frac{\angle \nabla_X(x),\nabla_{X'}(x)}{2} \leq 
\left( \frac{1}{1- \frac{ \sqrt{2}}{2}} \left(1 + \frac{1}{1- \frac{1}{2} \left(1 + \frac{ \sqrt{2}}{2}\right) } \right) +3 \right) \sqrt{\frac{\epsilon}{t}} \leq 30 \sqrt{\frac{\epsilon}{t}}.
$$
\end{proof}

\begin{lemma}\label{lemma-hausdorffdistance}
Let $K$ and $K'$ be two compact sets of $\R^d$ whose $\mu$-reaches are greater than $r$. 
Then for every 
$t\in (0, r\mu)$, we have:
$$
d_H(K_{r,t}, K'_{r,t}) \leq \frac{\epsilon}{\mu}
\quad \mbox{and}\quad
d_H(\partial K_{r,t},\partial K'_{r,t}) \leq \frac{\epsilon}{\mu}.
$$
\end{lemma}
\begin{proof}
First remark that if we take two compact sets $A$ and $B$  with $\tilde{\mu}$-reach greater than $r$, such that $d_H(A,B)\leq \tilde{\epsilon}$, one has:
\begin{equation} \label{eq:dist1}
d_H\left(
\overline{(A_r)^c},\overline{(B_r)^c}
\right)
\leq \frac{\tilde{\epsilon}}{\tilde \mu}
\end{equation}
Indeed, let $x \in \overline{A_r}^c$. Then $d(x,A)\geq r$ and $d(x,B)\geq r-\tilde{\epsilon}$. Let $s\mapsto \sigma(s)$, $\sigma(0)=x$ be the trajectory  of $\nabla_{B}$ issued from $x$ and parametrized by arc-length. While $\sigma(s)\in B_r$, we have \cite{Li}:
$$
d_{B}(\sigma(s)) = d_{B}(x) + \int_0^s \|\nabla_{B}(\sigma(s))\| ds \geq r-\tilde{\epsilon} + s\tilde{\mu}.
$$
We then have 
$\sigma(s) \in \overline{(B_r)^c}$ for $s\geq \frac{\tilde{\epsilon}}{\tilde{\mu}}$. As a consequence, there exists $x'\in \overline{(B_r)^c}$ such that $d(x,x')\leq \frac{\tilde{\epsilon}}{\tilde{\mu}}$. 
We apply Equation (\ref{eq:dist1}) with $A= K$ and $B=K'$ and we get:
\begin{equation} \label{eq:dist2}
d_H\left(
\overline{(K_r)^c},\overline{(K'_r)^c}
\right)
\leq \frac{{\epsilon}}{\mu}
\end{equation}
We apply again Equation (\ref{eq:dist1}) with $A= \overline{(K_r)^c}$ and $B=\overline{(K'_r)^c}$ with $\tilde{\epsilon}=\frac{\epsilon}{\mu}$ and $\tilde{\mu}=1$:
$$
d_H(\overline{{K_{r,t}}^c},
\overline{{K'_{r,t}}^c})
\leq \frac{\epsilon}{\mu}.
$$
Remark also that for any compact sets $A$ and $B$, one has $d_H(A_t,B_t)\leq d_H(A,B)$. Therefore, by Equation (\ref{eq:dist2}) one has:
$$
d_H({{K_{r,t}}},
{{K'_{r,t}}})
\leq \frac{\epsilon}{\mu}.
$$
The two last equations imply that
$$
d_H(\partial{{K_{r,t}}},
\partial{{K'_{r,t}}})
\leq \frac{\epsilon}{\mu}.
$$
Indeed, let $x \in \partial K_{r,t}$. Then $x$ is at a distance less than $\frac{\epsilon}{\mu}$ from $K'_{r,t}$ and $\overline{{K'_{r,t}}^c}$. Then, there exists $y\in K'_{r,t}$ and $z\in \overline{{K'_{r,t}}^c}$ such that $xy\leq \frac{\epsilon}{\mu}$ and $xz\leq \frac{\epsilon}{\mu}$. Since the line-segment $[yz]$ intersects $\partial K_{r,t}$, there exists $x'\in \partial K'_{r,t}$ such that $xx'\leq \frac{\epsilon}{\mu}$.
\end{proof}

\noindent{\bf Proof of Proposition \ref{proposition-double-offsets}}\\
Let $K$ and $K'$ be two compact sets whose $\mu$-reaches are greater than $r$. We put $\epsilon=d_H(K,K')$. We suppose that $\frac{\epsilon}{\mu\ (2-\sqrt{2})} \leq t\leq \frac{r\mu}{2}$. Lemma \ref{lemma-hausdorffdistance} then implies that 
$$
d_H(\partial K_{r,t},\partial K'_{r,t}) \leq \frac{\epsilon}{\mu} \leq   (2-\sqrt{2})\ t.
$$
Let $U_t$ be the tubular neighborhood of $\partial K_{r,t}$ of radius $t \leq reach(\partial K_{r,t})$. The projection map $\pi$ onto $\partial K_{r,t}$ is then defined on $U_t$. We clearly have $\partial K'_{r,t}\subset U_t$. More precisely, the map $\pi$ induces a one-to-one map between $\partial K'_{r,t}$ and $\partial K_{r,t}$ (see Theorem 4.1 of \cite{orthomap}).
We now define
$$
\begin{array}{cccc}
\psi:&U_t \times \R^d &\to & spt\left( N\left(K_{r,t}\right)\right)\\
&(x,n) &\mapsto &(\pi(x),n_{\pi(x)})
\end{array}.
$$
Let $h$ be the affine homotopy between $\psi$ and the identity (see Federer \cite{federer} 4.1.9 page 364)
$$
\begin{array}{cccc}
h:& [0,1] \times (U_t\times \R^d)&\to &\R^d\times \R^d\\
&(t,x)&\mapsto &(1-t)x+t\psi(x)\\
\end{array}.
$$

Since the map $\pi:\partial K'_{r,t}\to\partial K_{r,t}$ is one-to-one, the map $\psi$ also induces a one-to-one map between $spt(D')$ and $spt(D)$. Therefore, similarly as in the proof of Lemma \ref{lemma-erosions}, one has: 
$$
\psi_{\sharp}  D'=D.
$$
According to Federer (\cite{federer-59}, 4.1.9 page 363-364), we have:
$$
D-D'\ =\  \psi_{\sharp}  D' -  id_{\sharp}  D'\ = \partial R_1+R_2
$$
where $R_1 = \left[h_{\sharp}\left([0,1] \times D'\right)\right]$ and $R_2 = h_{\sharp}( [0,1]  \times \partial D')$. 
Again by Federer (\cite{federer-59}, 4.1.9 page 364) or Fanghua (\cite{fanghua} page 187), we get:
$$
M\left(h_{\sharp}\left([0,1] \times D'\right)\right)
\leq 
M\left(D' \right)\ 
\displaystyle \sup_{spt\left(D'\right)} |\psi-id|\
\displaystyle \sup_{spt\left(D'\right)} |1 + J_{d-1}(\psi)|,
$$
and
$$
M \left( h_{\sharp}( [0,1]  \times \partial D') \right)
\leq 
M\left(\partial D' \right)\ 
\displaystyle \sup_{spt\left(D'\right)} |\psi-id|\
\displaystyle \sup_{spt\left(D'\right)} |1 + J_{d-1}(\psi)|,
$$
where $J_{d-1}(\psi)$ stands for the (d-1)-dimensional jacobian (as defined in \cite{morgan}, page 24-25).
By Lemma \ref{lemma-hausdorffdistance}, the space component of $\psi-id$ is less than $\frac{\epsilon}{\mu}$. By Lemma \ref{lemma-normals}, the normal component of $\psi-id$ is less than $30\sqrt{\frac{\epsilon}{\mu t}}$. Thus
$$
\displaystyle \sup_{spt\left(N\left(K'_{r,t}\right)\right)} |\psi-id|
\leq \sqrt{\left(\frac{\epsilon}{\mu}\right)^2 + 900\frac{\epsilon}{\mu t}}.
$$
We can note that the jacobian of the space component of $\psi$ ({\it i.e.}  the projection) is bounded by $\left(\frac{t}{t-\frac{\epsilon}{\mu}}\right)^{d-1}$. 
The jacobian of the map $x\in K_{r,t} \mapsto n_x$ is upper bounded by  $\left(\frac{1}{t}\right)^{d-1}$ (where $t\leq \frac{\mu r}{2}$ is the reach of $\partial K_{r,t}$). The jacobian of the normal component of $\psi$ is upper bounded by the product $\left(\frac{1}{t}\ \frac{t}{t-\frac{\epsilon}{\mu}}\right)^{d-1}$. We then have
$$
\displaystyle \sup_{spt\left(N\left(K'_{r,t}\right)\right)} |1 + J_{d-1}(\psi)| 
\leq 1 + \left(\frac{1+t^2}{\left(t-\frac{\epsilon}{\mu}\right)^2}\right)^{\frac{d-1}{2}}.
$$
By using the definition of the flat norm we have:
$$
\F\left(D-D' \right)
\leq M
\left(h_{\sharp}\left([0,1] \times D'\right)\right)
+ M \left( h_{\sharp}( [0,1]  \times \partial D') \right).
$$
which is upper bounded by:
$$
\left( M\left(D'\right) + M\left(\partial D'\right) \right)
\left(1 + \left(\frac{1+t^2}{\left(t-\frac{\epsilon}{\mu}\right)^2}\right)^{\frac{d-1}{2}}\right)
\sqrt{\left(\frac{\epsilon}{\mu}\right)^2 + 900\frac{\epsilon}{\mu t}}.
$$

\section{From double offset to offset}\label{section-double-offset-2-offset}
The aim of that section is to prove Theorem \ref{thm:main}. We assume that $f$ is $C^1$, the general case being obtained by a limit argument. Let $C = (spt(f)_{\alpha}\times \R^d) \cap spt(N(K'_r))$, where $\alpha$ is a positive real number that will be specified later and let $B = F_t\circ i (C)$. 
\noindent
Thanks to Lemma \ref{lemma-erosions}, we have:
$$
\begin{array}{rl}
N\left(K'_r\right)\llcorner (i\circ F_{-t}(B))
&=-i_{\sharp}F_{-t \sharp} \left[N\left(K'_{r,t}\right)\llcorner B\right]
\end{array}.
$$
and
$$
\begin{array}{rl}
N\left(K_r\right)\llcorner (i \circ F_{-t} (\psi(B)))
&=-i_{\sharp}F_{-t \sharp}  \left[N\left(K_{r,t}\right) \llcorner \psi(B)\right].
\end{array},
$$
where $\psi$ is defined in the proof of Proposition  \ref{proposition-double-offsets}.
Remark that we have $N\left(K'_{r,t}\right)\llcorner B= N\left(K'_{r,t}\right)\llcorner (\tilde{B}\times \R^d)$,  and $N\left(K_{r,t}\right)\llcorner \psi(B)= N\left(K_{r,t}\right)\llcorner (\pi(\tilde{B})\times \R^d)$ where $\tilde{B}$ is the projection of $B$ onto the first factor of $\R^3\times \R^3$ and $\pi$ is the projection  onto $\partial K_{r,t}$.
Hence, Proposition \ref{proposition-double-offsets} allows us to write:
\begin{equation*}\label{eqn:}
N\left(K'_r\right)\llcorner (i\circ F_{-t}(B)) - N\left(K_r\right)\llcorner (i \circ F_{-t} (\psi(B))) = -i_\sharp F_{-t\sharp} (\partial R_1 + R_2),
\end{equation*}
where $R_1$ and $R_2$ satisfy certain properties. In particular, from the proof we see that $R_2= h_\sharp [(N(K'_{r,t})\llcorner \partial B) \times [0,1]]$, where $h$ is the result of the linear interpolation between the identity and $\psi$. Since $i\circ F_{-t}$ is linear, we have
$$
(i\circ F_{-t})\circ h = h' \circ \left(i\circ F_{-t},Id_{[0,1]}\right),
$$
where $h'$ is the result of the linear interpolation between the identity and $\psi' = i\circ F_{-t}\circ \psi \circ F_{t}\circ i$. We then have 
 that $-i_\sharp F_{-t\sharp}(R_2) = h'_\sharp [(-i_\sharp F_{-t\sharp}N(K'_{r,t})\llcorner \partial B) \times [0,1]] $.  Noting that $-i_\sharp F_{-t\sharp}(N(K'_{r,t})\llcorner \partial B) = N(K'_r)\llcorner \partial C$, and choosing $\alpha>||\psi'-Id||_\infty$, we see that the support of $i_\sharp F_{-t\sharp} R_2$ is included in the $\alpha$-offset of $spt(N(K'_r)\llcorner \partial C)$. Due to the choice of $\alpha$, this subset does not meet $spt(f)\times \R^d$. As a consequence, for any differential form $\omega$, letting:
$$
\Delta = |<N\left(K'_r\right)\llcorner (i\circ F_{-t}(B)),\bar{f}\omega> - <N\left(K_r\right) \llcorner (i \circ F_{-t} (\psi(B))),\bar{f}\omega>|, 
$$
we have:
\begin{eqnarray*}
 \Delta&\leq& |<i_\sharp F_{-t\sharp} \partial R_1,\bar{f}\omega>|\\
 &\leq& |<i_\sharp F_{-t\sharp} R_1,d\bar{f}\wedge\omega+\bar{f}d\omega>|\\
 &\leq& M(i_\sharp F_{-t\sharp} R_1)(Lip(\bar{f}) ||\omega||+||d \omega||)\\
 &\leq& (1+t^2)^{d/2}M(R_1)(Lip(f)||\omega||+||d \omega||)\\
 &\leq& M(N(K'_{r,t})\llcorner B) k_1k_2(1+t^2)^{d/2}(Lip(f)||\omega||+||d \omega||),
 \end{eqnarray*}
the last two inequalities following from Lemma \ref{lemma-erosions}, Proposition \ref{proposition-double-offsets}, and the fact that $Lip(f)=Lip(\bar{f})$.
Now remark that $i\circ F_{-t}(B) = C$ and $i\circ F_{-t}(\psi(B)) = \psi'(C)$ respectively contain $spt(N(K'_r))\cap(spt(f)\times \R^{d})$ and $spt(N(K_r))\cap(spt(f)\times \R^{d})$, due to the choice of $\alpha$. Hence we have:
$$
\Delta = |<N\left(K'_r\right),\bar{f}\omega> - <N\left(K_r\right),\bar{f}\omega>| .
$$
The quantity $M(N(K'_{r,t})\llcorner B)$ may be bounded above using Lemma \ref{lemma-mass-surface-c1}:
\begin{eqnarray*}
M(N(K'_{r,t})\llcorner B) &\leq& (1+reach(\partial K'_{r,t})^{-2})^{\frac{d-1}{2}}{\cal H}^{d-1}((\partial K'_{r,t})\cap B'),
\end{eqnarray*}
where $B'$ is the preimage of $spt(f)_{\alpha}\cap K'_r$ under the projection on $K'_r$. Since the reach of $\partial K'_{r,t}$ may be bounded by a function of $r$ and $\mu$ for $t=r\mu/2$, it only remains to bound the area term to finish the proof of Theorem \ref{thm:main}.  To do so, we use (a local version of) Theorem II.3 in \cite{boundarymeasure}, which yields the following bound in terms of covering numbers:
$$
{\cal H}^{d-1}((\partial K'_{r,t})\cap B') \leq {\cal N}(spt(f)_{\alpha}\cap \partial K'_r,t)\ \omega_{d-1}(2t)
$$
where $\omega_{d-1}(2t)$ denotes the surface area of the ball of radius $2t$ in $\R^d$. Note now that ${\cal N}(spt(f)_{\alpha}\cap \partial K'_r,t) \leq {\cal N}(spt(f)_{\alpha},t) $.
The fact that $\alpha = O(\sqrt \epsilon)$ easily follows from the following lemma:

\begin{lemma}\label{lemma-psi}
$$
\|\psi'-Id\|_{\infty} =O(\sqrt{\epsilon}).
$$
\end{lemma}
\begin{proof}
Let $(y',n') \in C$ and $x' \in \partial K'_{r,t}$ be such that the projection of $x'$ onto $\overline{{K'}_r^c}$ is $y'$. We put $x=\pi(x')$ the projection of  $x'$ onto $\partial K_{r,t}$ and we denote by $n$ the normal to $K_{r,t}$ at the point $x$ that is pointing inside.
We denote by $y$ the projection of $x$ onto $\overline{K_r^c}$. We then have:
$$
(x',-n')=F_t\circ i(y',n'),\quad
(x,-n)=\psi(x',-n')\quad
\mbox{and}\quad
(y,n)=i\circ F_{-t}(x,n)=\psi'(y',n').
$$
We now need to bound $yy'$ and $\|n-n'\|$. Lemma \ref{lemma-normals} implies that $\|n-n'\| =0(\sqrt{\epsilon})$.
We introduce $\tilde{y}$ the projection of $x'$ onto $\overline{K_r^c}$. Let us now bound  $y'\tilde{y}$ and $\tilde{y} y$. 

\begin{figure}[!h]
\psfrag{xp}{$x'$}
\psfrag{t}{$t$}
\psfrag{delta}{$\delta$}
\psfrag{yt}{$\tilde y$}
\psfrag{ys}{$y''$}
\psfrag{yp}{$y'$}
\psfrag{sb}{$\leq \frac{\e}{\mu}$}
\centerline{\includegraphics[height=5cm]{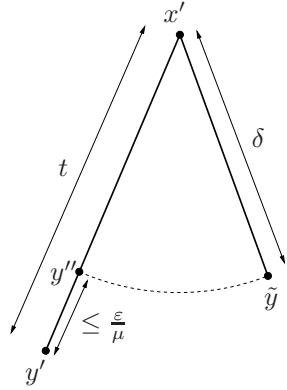}}
\caption{Proof of Lemma \ref{lemma-psi} }\label{fig:bound_phi}
\end{figure}

Since $x' \in \partial K'_{r,t}$, we have $x'y'=t$. Furthermore, by using Lemma \ref{lemma-hausdorffdistance}, one has 
$\delta=x'\tilde{y} \in \left[ t-\frac{\epsilon}{\mu},t+\frac{\epsilon}{\mu}\right]$. 
Let now denote by $y''$ the point of the half line issued from $x'$ and passing through $y'$, that is at a distance $\delta$ from $x'$. We have by Lemma \ref{lemma-normals}:
$$
\begin{array}{rl}
y'\tilde{y} 
&\leq y'y'' +y''\tilde{y}\\
&\leq \frac{\epsilon}{\mu} +  2 \sin \frac{\angle\left(\nabla_{\overline{K_r}}(x'),\nabla_{\overline{{K'_r}^c}}(x')\right)}{2}\ \left(t+\frac{\epsilon}{\mu}\right)\\
&\leq \frac{\epsilon}{\mu} + 30\ \sqrt{\frac{\epsilon}{\mu t}}\ \left(t+\frac{\epsilon}{\mu}\right)\\
&=0\left(\sqrt{ \epsilon}\right).
\end{array}
$$
Now by using again Lemma \ref{lemma-hausdorffdistance}, one has $xx' \leq \frac{\epsilon}{\mu}$ and also that $x$ and $x'$ are at a distance less than $\left(t+\frac{\epsilon}{\mu}\right)$ from $\overline{K_r^c}$. Since $\overline{K_r^c}$ has a reach greater than $\mu r$, then  the projection map onto $\overline{K_r^c}$ is $\frac{\mu r}{\mu r-\left(t+\frac{\epsilon}{\mu}\right)}$-Lipschitz for points at a distance less than $\left(t+\frac{\epsilon}{\mu}\right)$ from $\overline{K_r^c}$ (\cite{federer-59} page 435). We then have:
$$
\tilde{y} y \leq \frac{\epsilon}{\mu}\ \frac{\mu r}{\mu r-\left(t+\frac{\epsilon}{\mu}\right)}.
$$
We then clearly have $yy' \leq y'\tilde{y} + \tilde{y} y = 0(\sqrt{\epsilon})$.
\end{proof}

\section{Aknowledgements} 
The authors would like to thank S. Loriot who provided us with a software for computing the boundary of a union of balls.

\section{Conclusion and future works}
We have introduced the first  notion of anisotropic curvature measure which is Hausdorff stable and applies to a large class of objects,
 including non manifold and non smooth sets as well as point clouds.
Indeed, it is enough to require that some offset has  a positive $\mu$-reach, or, equivalently that the critical function of the set is
greater than some positive number $\mu$ on some interval.

In light of these results, one can introduce a scale dependent variant of the normal cycle. We say that a compact set $K \subset \R^d$ satisfies the $(P_\alpha)$ property if its critical function is greater than some positive 
$\mu$ on an open interval containing $\alpha>0$. For such $K$, we define the $\alpha$-normal cycle as the rectifiable $(d-1)$-current  of  $\R^d \times \S^{d-1}$, $N_{\alpha}(K) = F_{-\alpha \sharp} N(K_{\alpha})$. The effect of push-forward $ F_{-\alpha \sharp} $  is to move the support of the normal 
cycle closer to $K$: in simple cases (but not in general) $N_\alpha(K)$ is equal to the normal cycle of the double offset $\overline{{K_{\alpha,\alpha}}^c}$.
This current captures in some sense the curvature information at scale $\alpha$ and has two nice properties. 
First, it coincides with the usual normal cycle for sets with positive reach, more precisely, if a compact set $K$ has a reach  greater than $\alpha$, then $N_{\alpha}(K) = N(K)$. Second it is Hausdorff stable, more precisely,
if $K$ satisfies  $(P_\alpha)$, then there are constants $C$ and $\epsilon_0>0$ depending only on $K$ such that if $K'$ is a compact set such that
$d_H(K,K') < \epsilon \leq \epsilon_0$ then $N_{\alpha}(K)$ and $N_{\alpha}(K')$ differ by less than $C \sqrt \epsilon$ in the so-called
{\em flat norm} (see for example \cite{morgan,fanghua} for a definition), which implies that the associated curvature measures are also $O \left(\sqrt\epsilon \right)$ close.

We think of several possible future directions. First  we think of extending our paradigm to the measure of higher order quantities such as torsion of curves or curvature derivatives. A possible track to define a stable measure for these higher order quantities is to integrate the gradient  of a Gaussian function again 
the normal cycle: this would retrieve informations about the gradient of curvature measure.
Associated stability results require more investigations.

Furthermore, although the practical setting is beyond the scope of this paper, we obtain promising results for the estimation of the curvature measures from a noisy point cloud sample and we expect potential applications in the context of point cloud modeling.


\end{document}